\documentclass{aa501} 
\usepackage{psfig,lscape} 
 
\def\kms{$\rm km\;s^{-1}$} 
\def\kmsmpc{$\rm km\;s^{-1}\;Mpc^{-1}$} 
\def\kmspc{$\rm km\;s^{-1}\;pc^{-1}$} 
 
\def\dg{^\circ} 
\def\grad{$\Delta V / \Delta r$} 
\def\gradin{$(\Delta V / \Delta r)_{\it in}$} 
\def\gradout{$(\Delta V / \Delta r)_{\it out}$} 
\def\msun{M$_{\odot}$} 
 
\def\ha{H$\alpha$} 
\def\h2{H$_{2}$} 
 
\def\hii{H~{\small II}} 
\def\nii{[N~{\small II}]} 
 
\def\ng{[N~{\small II}]$\,\lambda6583$} 
 
\def\og{[O~{\small III}]$\,\lambda5007$}

\begin{document} 
 
\title{Position-velocity diagrams of ionized gas in the inner regions
  of disk galaxies\thanks{Based on observations carried out at
    European Southern Observatory (ESO N.58, A-0564), at the Multiple
    Mirror Telescope, which is a joint facility of the Smithsonian
    Institution and the University of Arizona, and at the Isaac Newton
    Telescope operated by the Isaac Newton group at the La Palma
    island at the Spanish Observatorio del Roque de los Muchachos of
    the Instituto de Astrofisica de Canarias.}}
 
\author{ 
     J.G.~Funes,~S.J.     \inst{1}, 
     E.M.~Corsini         \inst{2}, 
     M.~Cappellari \inst{3,}\thanks{ESA external fellow.}, 
     A.~Pizzella          \inst{4}, 
     J.C.~Vega Beltr\'an  \inst{5}, 
     C.~Scarlata          \inst{4,6}, 
     and F.~Bertola       \inst{4}}

\offprints{J.G. Funes, S.J.} 
\mail{jfunes@as.arizona.edu} 
 
\institute{ 
Vatican Observatory, University of Arizona, Tucson, AZ 85721, USA\and 
Osservatorio Astrofisico di Asiago, Dipartimento di Astronomia, 
  Universit\`a di Padova, via dell'Osservatorio~8, I-36012 Asiago, Italy\and 
Leiden Observatory, Postbus 9513, 2300 RA Leiden, The Netherlands \and 
Dipartimento di Astronomia, Universit\`a di Padova, 
  vicolo dell'Osservatorio~2, I-35122 Padova, Italy \and 
Instituto Astrof\'\i sico de Canarias, Calle Via Lactea s/n, 
  E-38200 La Laguna, Spain \and 
Space Telescope Science Institute, 3700 San Martin Dr., Baltimore, 
  MD 21218, USA}

\date{Received 11 December 2001 / Accepted 18 March 2002}

\titlerunning{Position-velocity diagrams of disk galaxies} 
\authorrunning{Funes et al.}

\abstract{We use long-slit spectroscopy along the major axis of a
  sample of 23 nearby disk galaxies to study the kinematic properties
  of the ionized-gas component in their inner regions. For each
  galaxy, we derive the position-velocity diagram of the ionized gas
  from its emission lines. We discuss the variety of shapes observed
  in such position-velocity diagrams by comparing the gas velocity
  gradient, velocity dispersion and integrated flux measured in the
  inner ($r\simeq\pm1''$) and outer regions ($r\simeq\pm4''$).
  This kind of analysis allows the identification of galaxies which
  are good candidates to host a circumnuclear Keplerian gaseous disk
  rotating around a central mass concentration, and to follow up with
  {\it Hubble Space Telescope\/} observations.
} 
 
\maketitle 
 
\section{Introduction} 
\label{sec:introduction} 
 
It is now commonly accepted that supermassive black holes (SMBHs
hereafter) are nearly ubiquitous in elliptical galaxies and bulges.
According to the standard paradigm, they should constitute the relics
of the intense quasar activity that occurred in the early phase of
galaxy evolution (see Ho 1999 for a review). The study of SMBHs has
greatly benefited from the high resolution capabilities of the
spectrographs onboard the {\it Hubble Space Telescope} (HST). Nowadays
about 40 galaxies, belonging to different morphological types are
known to harbor a SMBH. Their masses, $M_\bullet$, range from $10^6$
to $10^{10}$ \msun\ and have been derived from the analysis of the
stellar orbital structure or from the dynamics of gaseous
circumnuclear Keplerian disks (CNKD), using both optical and radio
observations (see Kormendy \& Gebhardt 2001 and Merritt \& Ferrarese
2001b for both the object list and a discussion of the accuracy of
$M_\bullet$ determinations).
 
The census of SMBHs is now large enough to probe the links between
$M_\bullet$ and the global properties of the host galaxies.
$M_\bullet$ correlates with the luminosity, $L_{\it sph}$, and
velocity dispersion, $\sigma$, of the spheroidal component of the host
galaxy (Kormendy \& Richstone 1995; Magorrian et al. 1998; Ferrarese
\& Merritt 2000; Gebhardt et al.  2000a) . A lively debate is ongoing
about the slope of the $M_\bullet-\sigma$ relation (Merritt \&
Ferrarese 2001a), although with its neglegible scatter it is a tighter
correlation than the $M_\bullet-L_{\it sph}$ relation.  The
consistency of $M_\bullet$ in active and quiescent galaxies has been
discussed by different authors.  Gebhardt et al.  (2000b) and
Ferrarese et al. (2001) showed that SMBH masses from reverberation
mapping agree with the $M_\bullet-\sigma$ relation.  Similarly, McLure
\& Dunlop (2001, 2002) found that the values of $M_\bullet$ inferred
from $L_{\it sph}$ for a large sample of quasars and Seyfert galaxies
agree with those inferred from $\sigma$ measurements.

\begin{landscape} 

\begin{table}[t] 
\caption{Parameters of the sample galaxies} 
\begin{footnotesize} 
\begin{center} 
\begin{tabular}{lllrrc rrr rrcrr} 
\hline 
\noalign{\smallskip} 
\multicolumn{1}{c}{Object} & 
\multicolumn{2}{c}{Morp. Type} & 
\multicolumn{1}{c}{$B_T$} & 
\multicolumn{1}{c}{P.A.} & 
\multicolumn{1}{c}{$i$} & 
\multicolumn{1}{c}{$V_{\odot}$} & 
\multicolumn{1}{c}{$D$} & 
\multicolumn{1}{c}{Scale} & 
\multicolumn{1}{c}{$M_{B_T}^0$} & 
\multicolumn{1}{c}{$\sigma_\star$} & 
\multicolumn{1}{c}{$M_\bullet$} & 
\multicolumn{1}{c}{$\theta_\bullet$} &  
\multicolumn{1}{c}{Nuc. Type} \\ 
\multicolumn{1}{c}{[name]} & 
\multicolumn{1}{c}{[RSA]} & 
\multicolumn{1}{c}{[RC3]} & 
\multicolumn{1}{c}{[mag]} & 
\multicolumn{1}{c}{[$\dg$]} & 
\multicolumn{1}{c}{[$\dg$]} & 
\multicolumn{1}{c}{[\kms]} & 
\multicolumn{1}{c}{[Mpc]} & 
\multicolumn{1}{c}{[pc/$''$]} & 
\multicolumn{1}{c}{[mag]} & 
\multicolumn{1}{c}{[\kms]} & 
\multicolumn{1}{c}{[\msun ]} & 
\multicolumn{1}{c}{[$''$]} & 
\multicolumn{1}{c}{} \\ 
\multicolumn{1}{c}{(1)} & 
\multicolumn{1}{c}{(2)} & 
\multicolumn{1}{c}{(3)} & 
\multicolumn{1}{c}{(4)} & 
\multicolumn{1}{c}{(5)} & 
\multicolumn{1}{c}{(6)} & 
\multicolumn{1}{c}{(7)} & 
\multicolumn{1}{c}{(8)} & 
\multicolumn{1}{c}{(9)} & 
\multicolumn{1}{c}{(10)} & 
\multicolumn{1}{c}{(11)} & 
\multicolumn{1}{c}{(12)} & 
\multicolumn{1}{c}{(13)} & 
\multicolumn{1}{c}{(14)} \\ 
\noalign{\smallskip} 
\hline 
\noalign{\smallskip} 
\object{NGC~470}  & Sbc(s)II.3 & .SAT3.. & 12.53 & 155 & 52 & 2370 & 33.8 &  
163.9 & $-20.66$ & $ 56$ & 3.2$\,$e$+05$ &  0.003 & ...    \\  
\object{NGC~772}  & Sb(rs)I    & .SAS3.. & 11.09 & 130 & 54 & 2470 & 35.6 &  
172.7 & $-22.21$ & $124$ & 1.4$\,$e$+07$ &   0.02 & ...    \\  
\object{NGC~949}  & Sc(s)III   & .SAT3*\$& 12.40 & 145 & 58 &  620 & 11.4 &   
55.2 & $-18.50$ & $ 32$ & 2.3$\,$e$+04$ &  0.002 & ...    \\  
\object{NGC~980}  & ...        & .L..... & 13.16 & 110 & 58 & 5765 & 80.1 &  
388.2 & $-22.95$ & $226$ & 2.3$\,$e$+08$ &   0.06 & ...    \\ 
\object{NGC~1160} & ...        & .S..6*. & 13.50 &  50 & 62 & 2510 & 36.6 &  
177.4 & $-21.01$ & $ 24$ & 5.9$\,$e$+04$ & 0.0003 & ...    \\ 
\object{NGC~2179} & Sa         & .SAS0.. & 13.22 & 170 & 47 & 2885 & 36.5 &  
177.0 & $-19.98$ & $166$ & 5.4$\,$e$+07$ &   0.05 & ...    \\ 
\object{NGC~2541} & Sc(s)III   & .SAS6.. & 12.26 & 165 & 61 &  565 &  8.7 &   
42.2 & $-18.13$ & $ 53$ & 2.5$\,$e$+05$ &   0.01 & T2/H:  \\  
\object{NGC~2683} & Sb(on edge)& .SAT3.. & 10.64 &  44 & 78 &  400 &  5.3 &   
25.6 & $-18.99$ & $ 83$ & 2.0$\,$e$+06$ &   0.06 & L2/S2  \\  
\object{NGC~2768} & S0$_{1/2}$(6)& .E.6.*. &10.84 & 95 & 59 & 1331 & 19.4 &   
94.1 & $-20.74$ & $205$ & 1.5$\,$e$+08$ &    0.2 & L2     \\  
\object{NGC~2815} & Sb(s)I-II  & .SBR3*. & 12.81 &  10 & 72 & 2541 & 30.5 &  
147.7 & $-21.00$ & $168$ & 5.7$\,$e$+07$ &   0.07 & L/S2   \\ 
\object{NGC~2841} & Sb         & .SAR3*. & 10.09 & 147 & 65 &  640 &  9.6 &   
46.4 & $-20.33$ & $197$ & 1.2$\,$e$+08$ &    0.3 & L2     \\  
\object{NGC~3031} & Sb(r)I-II  & .SAS2.. &  7.89 & 157 & 59 & $-50$&  1.5 &    
7.2 & $-18.46$ & $173$ & 6.6$\,$e$+07$ &    1.5 & S1.5   \\  
\object{NGC~3281} & Sa         & .SAS2P* & 12.70 & 140 & 61 & 3380 & 41.1 &  
199.5 & $-21.25$ & $172$ & 6.4$\,$e$+07$ &   0.05 & S2     \\ 
\object{NGC~3368} & Sab(s)II   & .SXT2.. & 10.11 &   5 & 47 &  865 &  9.7 &   
47.1 & $-20.14$ & $129$ & 1.6$\,$e$+07$ &    0.1 & L2     \\  
\object{NGC~3521} & Sb(s)II-III& .SXT4.. &  9.04 & 163 & 63 &  825 &  8.5 &   
41.1 & $-20.35$ & $145$ & 2.8$\,$e$+07$ &    0.2 & H/L2:  \\  
\object{NGC~3705} & Sab(r)I-II & .SXR2.. & 11.86 & 122 & 66 & 1000 & 11.4 &   
55.2 & $-19.03$ & $109$ & 7.4$\,$e$+06$ &   0.05 & T2     \\  
\object{NGC~3898} & SaI        & .SAS2.. & 11.60 & 107 & 54 & 1184 & 17.1 &   
82.9 & $-19.85$ & $223$ & 2.2$\,$e$+08$ &    0.3 & T2     \\  
\object{NGC~4419} & SBab:      & .SBS1./ & 12.08 & 133 & 71 &$-210$& 17.0 &   
82.4 & $-19.55$ & $ 98$ & 4.5$\,$e$+06$ &   0.03 & H      \\  
\object{NGC~4698} & Sa         & .SAS2.. & 11.46 & 170 & 52 &  992 & 17.0 &   
82.4 & $-19.91$ & $174$ & 6.7$\,$e$+07$ &    0.1 & S2     \\  
\object{NGC~5064} & Sa         & PSA.2*. & 13.04 &  38 & 63 & 2980 & 36.0 &  
174.4 & $-21.11$ & $202$ & 1.4$\,$e$+08$ &   0.09 & L      \\ 
\object{NGC~7320} & ...        & .SAS7.. & 13.23 & 132 & 60 &  862 & 15.4 &   
74.7 & $-18.39$ & ...   & ...           &    ... & ...    \\         
\object{NGC~7331} & Sb(rs)I-II & .SAS3.. & 10.35 & 171 & 70 &  820 & 14.7 &   
72.0 & $-21.48$ & $141$ & 2.5$\,$e$+07$ &   0.09 & T2     \\  
\object{NGC~7782} & Sb(s)I-II  & .SAS3.. & 13.08 & 175 & 58 & 5430 & 75.3 &  
364.9 & $-21.95$ & $193$ & 1.1$\,$e$+08$ &   0.04 & ...    \\ 
\noalign{\smallskip} 
\hline 
\noalign{\medskip} 
\end{tabular} 
\end{center} 
\begin{minipage}{24cm} 
NOTES -- Col.(2): morphological classification from RSA and RC3. 
Col.(3): morphological classification from RC3. 
Col.(4): total observed blue magnitude from RC3 except for 
         NGC~980 and NGC 5064 (LEDA). 
Col.(4): major-axis position angle taken from RC3. 
Col.(6): inclination derived from $\cos^{2}{i}\,=\,(q^2-q_0^2)/(1-q_0^2)$. 
         The observed axial ratio $q$ is taken from RC3 and the 
         intrinsic flattening $q_0=0.11$ has been assumed following 
         Guthrie (1992). 
Col.(7): heliocentric velocity of the galaxy derived from the center of 
         symmetry of the rotation curve of the gas. They are taken from Vega 
         Beltr\'an et al. (2001) except for NGC~2179, NGC~3281 
         and NGC~4698 (Corsini et al. 1999). 
Col.(8): distance obtained as $V_0/H_0$ with $H_0=75$ \kmsmpc\ 
         and $V_0$ the systemic velocity derived 
         from $V_\odot$ corrected for the motion of the Sun with 
         respect of the Local Group according to the RSA. 
         For NGC~4419 and NGC~4698, which belong to the Virgo 
         cluster, we assume a distance of 17 Mpc (Freedman et 
         al. 1994). 
Col.(10): absolute total blue magnitude corrected for 
         inclination and extinction from RC3. 
Col.(11): central velocity dispersion of the stellar component. Data are
         from Vega Beltr\'an et al. (2001) except for NGC 2179, 
         NGC 3281 (Corsini et al. 1999), NGC 2768 (Bertola et al. 
         1995), NGC 2815 (LEDA), NGC 3521 (Zeilinger et al. 2001), 
         and NGC 4698 (Bertola et al. 1999).
Col.(12): expected mass of the central SMBH derived from
          $\sigma_\star$ following Merritt \& Ferrarese (2001a).  
Col.(13): angular size of the radius of influence of the central 
          SMBH derived from $\theta_\bullet \approx
          M_\bullet\;\sigma_{\star}^{-2}\;D^{-1}$ (de Zeeuw 2001), where
          $M_\bullet$ is the mass of the SMBH in units of $10^6$ \msun ,
          $\sigma_\star$ is the central velocity dispersion in units of 100
          \kms , and $D$ is the galaxy distance in Mpc.       
Col.(14): classification of the nuclear spectrum from Ho,
        Filippenko \& Sargent (1997) except NGC 2815 (V\'eron-Cetty \&
        V\'eron 1986).  H = \hii\ nucleus. L = LINER. S = Seyfert
        nucleus.  T = transition object. ... = uncertain.
\end{minipage} 
\end{footnotesize} 
\label{tab:sample_properties} 
\end{table} 

\end{landscape}

Elliptical and disk galaxies seem to follow the same 
$M_\bullet-L_{\it sph}$ and $M_\bullet-\sigma$ correlations, 
suggesting a close connection between the processes leading to the
growth of central SMBHs and the formation of galaxy spheroids, whether
they are ellipticals, classical bulges or pseudobulges (Kormendy
2001). Morever, $M_\bullet$ does not correlate with disks as it does
with spheroids. To date, however, dynamical SMBH detections are
available for only a dozen disk galaxies, and, therefore, the addition
of new $M_\bullet$ determinations for S0's and spirals is highly
desirable.
 
Over the course of the last few years, we have undertaken a vast
program aimed at detecting CNKDs in disk galaxies using ground-based
spectroscopic observations. Our goal is to measure upper limits for
SMBH masses by using HST spectra to constrain them further (Bertola et
al. 1998).
Here we present a survey of the ionized-gas kinematics of the inner
regions of 23 disk galaxies. The study complements the recent results
regarding rapidly rotating gaseous nuclear disks in Rubin,
Kenney \& Young (1997) and Sofue et al. (1998).
The paper is organized as follows: in Sect. \ref{sec:observations} we
give an overview of the global properties of the sample galaxies and
discuss observations and data reduction. In Sect.
\ref{sec:PV_diagrams} we derive the position-velocity (PV) diagrams of
the emission lines for each sample galaxy, suggesting a scheme for
their classification. In Sect. \ref{sec:conclusions} we present our
conclusions. Relevant properties and the PV diagram of individual
galaxies are discussed in the appendix.

\section{Sample selection, spectroscopic observations and data reduction} 
\label{sec:observations} 
 
\begin{figure} 
\centerline{{\psfig{figure=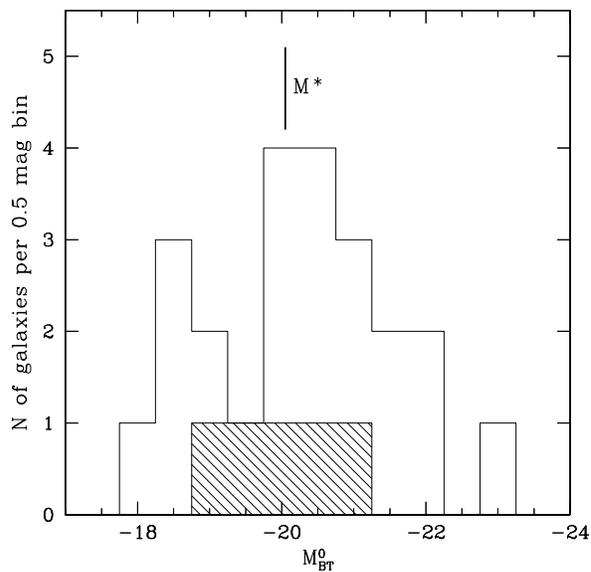,width=8cm}}} 
\caption{Absolute magnitude distribution for the sample galaxies. 
  A line marks $M_{B_T}^0 = -20.05$, which corresponds to M$^\ast$ for
  spiral galaxies as derived by Marzke et al. (1998) and assuming $H_0
  = 75$ \kmsmpc . The dashed region identifies galaxies classified
  barred or weakly barred in RC3.}
\label{fig:histogram} 
\end{figure}

All the observed galaxies are bright ($B_T\leq13.5$) and nearby
($V_\odot < 5800$ \kms) with an intermediate-to-high inclination
($45\dg \leq i \leq 80\dg$). The Hubble morphological types of the
sample galaxies range from S0 to Sd and 5 objects are barred or weakly
barred (RC3).  The galaxies have been chosen to have strong emission
lines. An overview of their basic properties is given in
Tab.~\ref{tab:sample_properties}.  In Fig. \ref{fig:histogram} we show
the absolute magnitude distribution for the galaxies in our sample.
The distribution brackets the M$^\ast$ value for spiral galaxies
taken from Marzke et al. (1998) for $H_0 = 75$ \kmsmpc .

The long-slit spectroscopic observations of our sample galaxies were 
carried out at the 4.5-m Multiple Mirror Telescope (MMT) in Arizona 
(USA), at the 3.6-m ESO Telescope at La Silla (Chile), and at the 
2.5-m Isaac Newton Telescope (INT) at La Palma (Spain). Details about 
the instrumental setup and the seeing of each observing run are  
summarized in Tab. \ref{tab:setup}. 
 
At the beginning of each exposure, the slit was centered on the galaxy
nucleus and aligned along the galaxy major axis using the guide TV
camera. The slit orientation and the exposure times are given in Tab.
\ref{tab:log_spectroscopy}. A comparison-lamp spectrum was obtained
before and after each object integration to allow an accurate
wavelength calibration. Quartz-lamp and twilight-sky flatfields were
taken to map pixel-to-pixel sensitivity variations and large-scale
illumination patterns.
  
Using standard MIDAS\footnote{MIDAS is developed and maintained by the
  European Southern Observatory.} routines, the spectra were bias
subtracted, flatfield corrected, cleaned for cosmic rays, and
wavelength calibrated. Cosmic rays were identified by comparing the
counts in each pixel with the local mean and standard deviation (based
on the Poisson statistics of the photons are using the gain and
readout noise of the detector). We corrected the data by interpolating
a suitable value.
 
The instrumental resolution was derived as the mean of the Gaussian
FWHMs measured for a number of unblended arc-lamp lines (12 in the MMT
and INT spectra and 30 in the ESO spectra), which were distributed
over the whole spectral range of a wavelength-calibrated comparison
spectrum. The mean FWHM of the arc-lamp lines, as well as the
corresponding instrumental velocity dispersion, are given in Tab.
\ref{tab:setup}. Finally, the spectra were aligned and coadded using
the centers of their stellar continua as reference. In the resulting
spectra, the contribution from the sky was determined by interpolating
along the outermost $10''-20''$ at the edges of the slit, where galaxy
light was negligible. The sky level was then subtracted.

\begin{table*}[ht!] 
\caption{Instrumental setup for spectroscopic observations} 
\begin{center} 
\begin{tabular}{lccccc} 
\hline 
\noalign{\smallskip} 
\multicolumn{1}{c}{Parameter} & 
\multicolumn{2}{c}{MMT} & 
\multicolumn{1}{c}{INT} & 
\multicolumn{1}{c}{3.6-m ESO} \\ 
\noalign{\smallskip} 
\hline 
\noalign{\smallskip} 
Date                  & 21--23 Oct 1990 & 17--18 Dec 1990       & 19--20 Mar  
1996 & 03--04 Feb 1997  \\ 
Spectrograph          & \multicolumn{2}{c}{Red Channel$\rm ^a$} & IDS$\rm ^a$ &  
CASPEC$\rm ^b$  \\ 
Grating ($\rm grooves\;mm^{-1}$) & \multicolumn{2}{c}{1200}     &  1800         
& 31.6\\ 
Detector              & \multicolumn{2}{c}{Loral 12$\times$8mmt}& TK1024A        
 & TK1024AB\\ 
Pixel size ($\rm \mu m^{2}$) & \multicolumn{2}{c}{$15\times15$} & $24\times24$   
  & $24\times24$\\ 
Pixel binning         &  \multicolumn{2}{c}{$1\times1$}         & $1\times1$     
  & $1\times1$\\ 
Scale ($\rm ''\;pixel^{-1}$)  & \multicolumn{2}{c}{0.30}        &  0.33          
  & 0.33\\ 
Reciprocal dispersion ($\rm \AA\;pixel^{-1}$) & \multicolumn{2}{c}{0.82} & 0.24  
   & 0.076\\ 
Slit width ($''$)   & \multicolumn{2}{c}{1.25}                  &   1.9          
  & 1.3\\ 
Slit length ($'$)   & \multicolumn{2}{c}{3.0}                   &   4.0          
  & 2.4\\ 
Spectral range (\AA)     & \multicolumn{2}{c}{4850--5500}       & 6650--6890     
  & 6630--6651\\ 
Comparison lamp      & \multicolumn{2}{c}{He--Ne--Ar--Fe}       & Cu--Ar         
  & Th--Ar\\ 
Instrumental FWHM (\AA) & $2.24\pm0.26$  & $2.57\pm0.11$        &  
$0.869\pm0.040$ & $0.233\pm0.017$ \\ 
Instrumental $\sigma$$\rm ^c$ (\kms) & 57 & 65 & 5 & 17 \\ 
Seeing FWHM ($''$)           & \multicolumn{2}{c}{1.2--1.5}      & 1.0--1.4      
   & 0.8--1.2\\ 
\noalign{\smallskip} 
\hline 
\noalign{\medskip} 
\end{tabular} 
\end{center} 
\begin{minipage}{18cm} 
$\rm ^a$ Grating used at the first order.\\ 
$\rm ^b$ CASPEC mounting of the Long Camera in long-slit configuration 
         without crossdisperser. The spectral order \#86 
         ($\lambda_c = 6617$ \AA) corresponding to the redshifted 
         \ha\ 
         region was isolated by means of the narrow-band 6630/51 \AA\ 
         filter. \\ 
$\rm ^c$ The instrumental dispersion has been measured at \og , 
         \ng\ and \ha\ for the MMT, INT and 3.6-m ESO spectra, 
         respectively. 
\end{minipage} 
\label{tab:setup} 
\end{table*}

\begin{table}[ht!] 
\caption{Log of spectroscopic observations} 
\begin{tabular}{lrlrr} 
\hline 
\noalign{\smallskip} 
\multicolumn{1}{c}{Object} & 
\multicolumn{1}{c}{Date}& 
\multicolumn{1}{c}{Telescope}& 
\multicolumn{1}{c}{t$_{\it exp}$} & 
\multicolumn{1}{c}{P.A.}\\ 
\noalign{\smallskip} 
\multicolumn{1}{c}{[name]} & 
\multicolumn{1}{c}{} & 
\multicolumn{1}{c}{} & 
\multicolumn{1}{c}{[s]} & 
\multicolumn{1}{c}{[$\dg$]} \\ 
\noalign{\smallskip} 
\hline 
\noalign{\smallskip} 
NGC~470  & 22 Oct 90 & MMT       & 3600          & 155 \\ 
NGC~772  & 22 Oct 90 & MMT       & 3600          & 130 \\ 
NGC~949  & 21 Oct 90 & MMT       & 3600          & 145 \\ 
NGC~980  & 22 Oct 90 & MMT       & 3600          & 110 \\ 
NGC~1160 & 21 Oct 90 & MMT       & 3600          &  50 \\ 
NGC~2179 & 03 Feb 97 & 3.6-m ESO & 4$\times$3600 & 170 \\ 
         & 04 Feb 97 & 3.6-m ESO & 2$\times$3600 & 170 \\ 
NGC~2541 & 21 Oct 90 & MMT       & 3600          & 165 \\ 
NGC~2683 & 18 Dec 90 & MMT       & 3600          &  44 \\ 
NGC~2768 & 19 Mar 96 & INT       & 2$\times$3600 &  95 \\ 
NGC~2815 & 04 Feb 97 & 3.6-m ESO & 3$\times$3600 &  10 \\ 
NGC~2841 & 22 Oct 90 & MMT       & 3600          & 147 \\ 
NGC~3031 & 17 Dec 90 & MMT       & 3600          & 157 \\ 
NGC~3281 & 04 Feb 97 & 3.6-m ESO & 3600          & 140 \\ 
NGC~3368 & 17 Dec 90 & MMT       & 3600          &   5 \\ 
NGC~3521 & 17 Dec 90 & MMT       & 3600          & 342 \\ 
NGC~3705 & 17 Dec 90 & MMT       & 3600          & 122 \\ 
NGC~3898 & 19 Mar 96 & INT       & 3$\times$3600 & 107 \\ 
NGC~4419 & 20 Mar 96 & INT       & 2$\times$3300 & 133 \\ 
         & 20 Mar 96 & INT       & 3600          & 133 \\ 
NGC~4698 & 20 Mar 96 & INT       & 3600          & 170 \\ 
NGC~5064 & 03 Feb 97 & 3.6-m ESO & 3$\times$3600 &  38 \\ 
         & 04 Feb 97 & 3.6-m ESO & 3600          &  38 \\ 
NGC~7320 & 22 Oct 90 & MMT       & 3600          & 132 \\ 
NGC~7331 & 22 Oct 90 & MMT       & 3600          & 171 \\ 
NGC~7782 & 22 Oct 90 & MMT       & 3600          &  30 \\ 
\noalign{\smallskip} 
\hline 
\end{tabular} 
\label{tab:log_spectroscopy} 
\end{table}

\section{Position-velocity diagrams} 
\label{sec:PV_diagrams} 
 
\subsection{Galaxy continuum subtraction} 
\label{sec:continuum} 
 
We subtracted the stellar continua from the spectra to study the
two-dimensional shape of the emission lines. Without a large library
of stellar (or even galaxy) absorption-line spectra obtained with the
same observing setup of the spectra of the sample galaxies we were
unable to apply the technique of template subtraction for an optimal
correction for the starlight contamination (e.g. Ho et al. 1997).
 
The galaxy continua have been removed from MMT spectra by applying the
technique outlined by Bender, Saglia \& Gerhard (1994). The following
procedure was applied to the each row of the galaxy spectrum. First we
fit a sixth-to-tenth-order polynomial to the observed spectrum and
calculated the rms variation, $\sigma$, of the spectrum around the
polynomial. Then, the fit was repeated including only those pixels
with values falling within $-1\sigma$ to $0\sigma$ of the first fit in
order to avoid both emission and strong absorption lines. The new
polynomial fit was adopted as the galaxy continuum and subtracted from
the observed spectrum.
We were prevented from adopting the same method for the INT and ESO
spectra because of their shorter wavelength range. Our major concern,
with the stellar continuum subtraction was avoiding the creation of
spurious features. For this reason we adopted a very simple but robust
approach. Specifically, we made the reasonable assumption that the
underlying observed stellar profile is the same at all  
wavelengths in the small observed range. An average profile was
determined in regions free from emission-line flux and this same
profile, properly scaled and subtracted from all the columns of
the spectrum. The stellar continuum under the emission features
was approximated by linear interpolation.
 
\begin{figure*} 
\psfig{figure=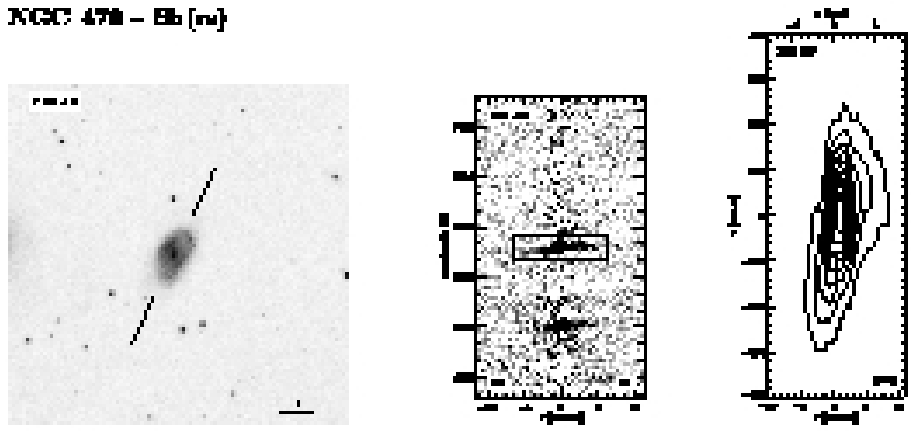,height=5.8cm} 
\vspace{0.2cm} 
\psfig{figure=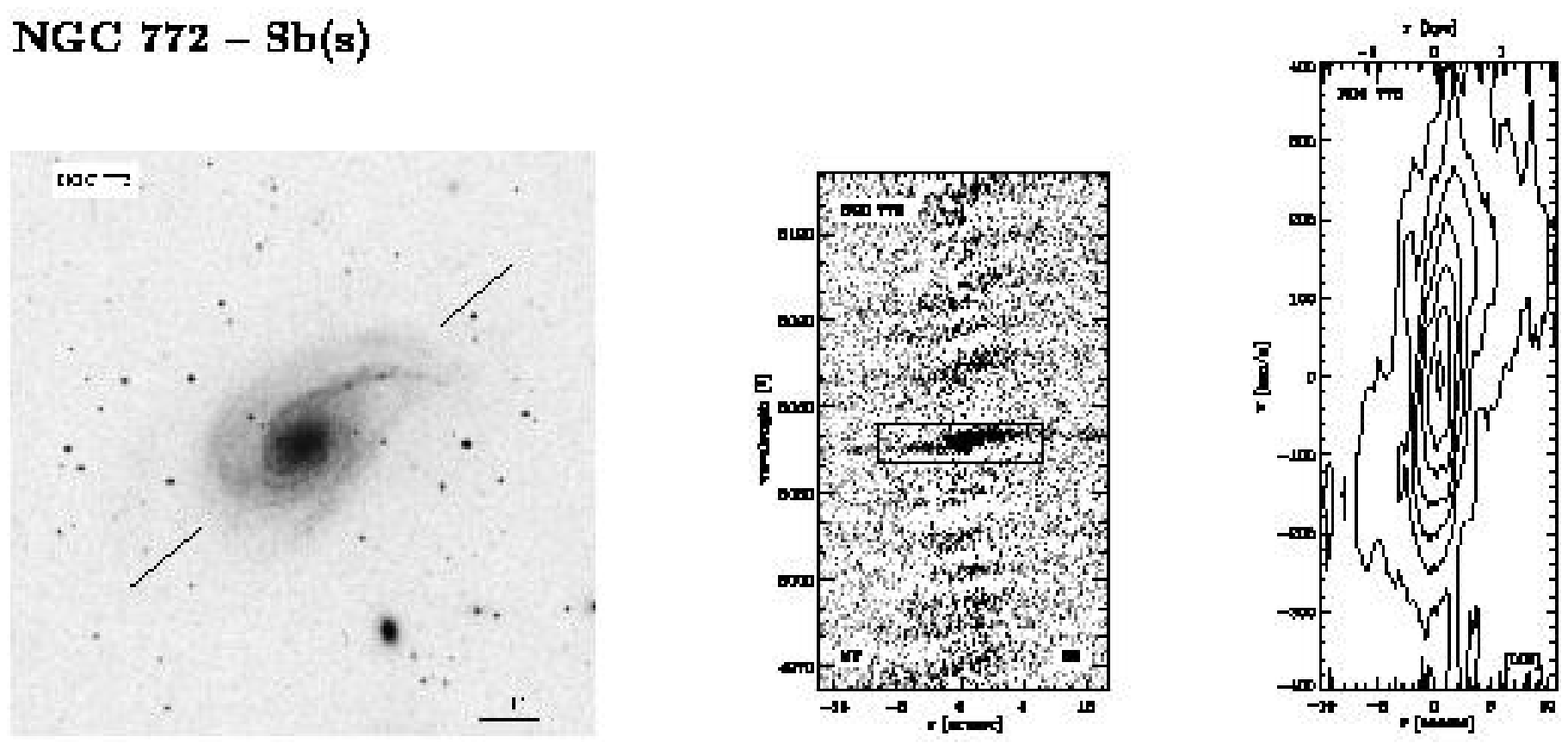,height=5.8cm} 
\bigskip 
\bigskip 
\bigskip 
\caption{Optical images, spectra and PV 
  diagrams of the sample galaxies.  We show from left to right: 
  a) An optical image of the galaxy taken from the Digitized Sky 
  Survey with the slit position and angular scale superimposed. The 
  orientation of the image is north up and east to the left. 
  b) The galaxy spectrum after continuum removal with wavelength, 
  radial distance from the nucleus, and orientation marked. 
  Color cuts are chosen to show the fainter structures and the radial 
  extension of the emission lines. In the INT and ESO spectra the 
  nuclear continuum is the residual after subtraction of about $90\%$ 
  of the continuum. 
  c) The galaxy PV diagram derived from \og, \ng, and the \ha\ 
  emission line for the spectra obtained at the MMT, INT, and 3.6-m ESO  
  telescopes, respectively. In the PV diagram the intensities of the 
  plotted contours correspond to $5\%, 15\%, 25\%,$ \ldots, $95\%$ 
  of the maximum emission-line intensity.  The plotted region in the 
  PV diagram corresponds to the rectangular box marked in the galaxy 
  spectrum. PV diagrams are shown with the same scale for the observed 
  radii and velocities, but we also indicate the distance from the 
  center in kpc to aid comparison.} 
\label{fig:atlas} 
\end{figure*}

For our purposes, the above techniques give a satisfactory
approximation of the galaxy continuum in the spectral range centered
on the relevant emission lines we measure, specifically \og , \ng, and
\ha\ for the MMT, INT and ESO spectra, respectively. In Fig.
\ref{fig:atlas} we show the continuum-subtracted spectra of the sample
galaxies as well as the isodensity contour plots (i.e. the PV diagram)
of the emission lines we measure.
 
\begin{table*}[ht!] 
\caption{Measured parameters of the sample galaxies} 
\begin{flushleft} 
\begin{footnotesize} 
\begin{center} 
\begin{tabular}{lllrrrcrc} 
\hline 
\noalign{\smallskip} 
\multicolumn{1}{c}{Object} & 
\multicolumn{1}{c}{$(\Delta V / \Delta r)_{\it in}$} & 
\multicolumn{1}{c}{$(\Delta V / \Delta r)_{\it out}$} & 
\multicolumn{1}{c}{$\Gamma$} & 
\multicolumn{1}{c}{$\sigma_0$} & 
\multicolumn{1}{c}{$\sigma_{\it out}$} & 
\multicolumn{1}{c}{$\sigma_0/\sigma_{\it out}$} & 
\multicolumn{1}{c}{$F_{0}/F_{\it out}$} & 
\multicolumn{1}{c}{Type} \\ 
\multicolumn{1}{c}{[name]} & 
\multicolumn{1}{c}{[\kms\ arcsec$^{-1}$]} & 
\multicolumn{1}{c}{[\kms\ arcsec$^{-1}$]} & 
\multicolumn{1}{c}{} & 
\multicolumn{1}{c}{[\kms]} & 
\multicolumn{1}{c}{[\kms]} & 
\multicolumn{1}{c}{} & 
\multicolumn{1}{c}{} & 
\multicolumn{1}{c}{} \\ 
\multicolumn{1}{c}{(1)} & 
\multicolumn{1}{c}{(2)} & 
\multicolumn{1}{c}{(3)} & 
\multicolumn{1}{c}{(4)} & 
\multicolumn{1}{c}{(5)} & 
\multicolumn{1}{c}{(6)} & 
\multicolumn{1}{c}{(7)} & 
\multicolumn{1}{c}{(8)} & 
\multicolumn{1}{c}{(9)} \\
\noalign{\smallskip} 
\hline 
\noalign{\smallskip} 
NGC~470  & $33\pm 4$ &$23\pm 7$ &$1.4^{+0.9}_{-0.5}$ &$101\pm15$  
 & $103\pm41$ &$1.0^{+0.9}_{-0.4}$ & $4.0\pm0.1$ &III\\ 
NGC~772  & $32\pm 3$ &$33\pm19$ &$1.0^{+1.5}_{-0.4}$ &$131\pm19$  
 & $138\pm51$ &$0.9^{+0.8}_{-0.4}$ & $3.7\pm0.1$ &III\\ 
NGC~949  & $ 6\pm 3$ &$ 6\pm 4$ &$1.0^{+3.5}_{-0.7}$ &$ 57\pm13$  
 & $ 42\pm23$ &$1.4^{+2.3}_{-0.7}$ & $2.7\pm0.1$ &III\\ 
NGC~980  &$108\pm12$ &$52\pm11$ &$2.1^{+0.8}_{-0.6}$ &$259\pm34$  
 & $129\pm95$ &$2.0^{+6.6}_{-1.0}$ & $6.9\pm0.1$ &I\\ 
NGC~1160 & $17\pm14$ &$33\pm24$ &$0.5^{+2.9}_{-0.5}$ &$ 66\pm21$  
 & $ 32\pm11$ &$2.0^{+2.1}_{-1.0}$ & $1.2\pm0.1$ &II\\ 
NGC~2179 & $86\pm 4$ &$41\pm 6$ &$2.1^{+0.5}_{-0.4}$ &$170\pm29$  
 & $ 42\pm11$ &$4.0^{+2.4}_{-1.4}$ & $4.2\pm0.1$ &I\\ 
NGC~2541 & $ 7\pm 1$ &$ 4\pm 2$ &$1.8^{+2.3}_{-0.8}$ &$ 48\pm18$  
 & $ 55\pm13$ &$0.9^{+0.7}_{-0.4}$ & $5.2\pm0.2$ &III\\ 
NGC~2683 & $36\pm 6$ &$12\pm 8$ &$3.0^{+7.5}_{-1.5}$ &$109\pm13$  
 & $163\pm31$ &$0.7^{+0.3}_{-0.2}$ & $3.2\pm0.1$ &*\\ 
NGC~2768 & $20\pm 2$ &$ 7\pm 4$ &$2.9^{+4.5}_{-1.2}$ &$174\pm12$  
 & $141\pm48$ &$1.2^{+0.8}_{-0.4}$ &$10.8\pm0.1$ &III\\ 
NGC~2815 & $49\pm 1$ &$49\pm 6$ &$1.0^{+0.2}_{-0.1}$ &$149\pm16$  
 & $ 94\pm 9$ &$1.6^{+0.4}_{-0.3}$ &$15.0\pm0.5$ &III\\ 
NGC~2841 & $15\pm 7$ &$17\pm 2$ &$0.9^{+0.6}_{-0.5}$ &$135\pm18$  
 & $144\pm19$ &$0.9^{+0.3}_{-0.2}$ & $2.5\pm0.1$ &III\\ 
NGC~3031 & $21\pm 4$ &$ 7\pm 3$ &$3.0^{+3.3}_{-1.3}$ &$237\pm51$  
 & $ 57\pm41$ &$4.2^{+13.8}_{-2.3}$ &$65.6\pm0.4$ &III\\ 
NGC~3281 & $34\pm 5$ &$18\pm 9$ &$1.9^{+2.4}_{-0.8}$ &$119\pm20$  
 & $ 92\pm12$ &$1.3^{+0.4}_{-0.3}$ &$19.9\pm0.5$ &III\\ 
NGC~3368 & $35\pm12$ &$32\pm 8$ &$1.1^{+0.9}_{-0.5}$ &$113\pm19$  
 & $ 69\pm12$ &$1.6^{+0.7}_{-0.5}$ & $6.1\pm0.2$ &III\\ 
NGC~3521 & $21\pm10$ &$18\pm11$ &$1.2^{+3.3}_{-0.8}$ &$180\pm39$ 
 & $146\pm66$ &$1.2^{+1.5}_{-0.6}$ & $8.9\pm0.1$ &III\\ 
NGC~3705 & $22\pm 1$ &$14\pm10$ &$1.6^{+4.2}_{-0.7}$ &$110\pm14$  
 & $ 55\pm42$ &$2.0^{+7.5}_{-1.0}$ &$13.3\pm0.4$ &III\\ 
NGC~3898 & $13\pm 1$ &$26\pm 3$ &$0.5^{+0.1}_{-0.1}$ &$131\pm21$  
 & $110\pm27$ &$1.2^{+0.6}_{-0.4}$ & $4.6\pm0.1$ &III\\ 
NGC~4419 & $ 7\pm 1$ &$14\pm 2$ &$0.5^{+0.2}_{-0.1}$ &$117\pm11$  
 & $ 83\pm21$ &$1.4^{+0.7}_{-0.4}$ &$13.6\pm0.1$ &III\\ 
NGC~4698 & $ 8\pm 5$ &$ 3\pm 1$ &$2.7^{+3.8}_{-1.9}$ &$ 86\pm 8$  
 & $103\pm 7$ &$0.8^{+0.1}_{-0.1}$ & $6.0\pm0.1$ &III\\ 
NGC~5064 & $74\pm11$ &$56\pm11$ &$1.3^{+0.6}_{-0.4}$ &$ 52\pm22$  
 & $ 43\pm11$ &$1.2^{+1.1}_{-0.7}$ & $0.9\pm0.1$ &II\\ 
NGC~7320 & $ 6\pm 1$ &$ 5\pm 3$ &$1.2^{+2.3}_{-0.6}$ &$ 12\pm10$  
 & $ 10\pm7 $ &$1.2^{+6.1}_{-1.1}$ & $0.3\pm0.1$ &II\\ 
NGC~7331 & $16\pm 4$ &$15\pm11$ &$1.1^{+3.9}_{-0.6}$ &$102\pm11$  
 & $130\pm65$ &$0.8^{+1.0}_{-0.3}$ & $3.7\pm0.1$ &III\\ 
NGC~7782 &$170\pm25$ &$56\pm11$ &$3.0^{+1.3}_{-0.9}$ &$151\pm19$  
 & $ 96\pm29$ &$1.6^{+1.0}_{-0.5}$ & $5.3\pm0.1$ &I\\ 
\noalign{\smallskip} \hline 
\noalign{\smallskip} 
\noalign{\smallskip} 
\noalign{\smallskip} 
\end{tabular}
\begin{minipage}{18cm} 
NOTES -- 
Col.(2): inner velocity gradient at $r\simeq\pm1''$. 
Col.(3): outer velocity gradient at $r\simeq\pm4''$. 
Col.(4): inner-to-outer velocity gradient ratio. 
Col.(5): central velocity dispersion. 
Col.(6): outer velocity dispersion at $r\simeq\pm4''$. 
Col.(7): central-to-outer velocity dispersion ratio. 
Col.(8): Central-to-outer integrated-flux ratio. $F_0$ and 
         $F_{\it out}$ have been measured at $r=0''$ and 
         $r\simeq\pm4''$, respectively. 
Col.(9): type of PV diagram according to our classification; 
         * = figure-of-eight PV diagram (see appendix). 
\end{minipage} 
\end{center} 
\end{footnotesize} 
\end{flushleft} 
\label{tab:measured_properties} 
\end{table*} 
 
\addtocounter{figure}{-1} 
\begin{figure*} 
\psfig{figure=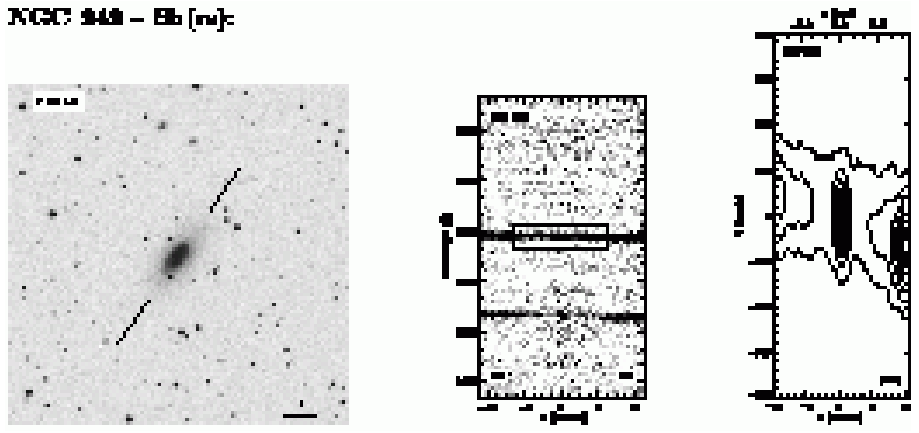,height=5.8cm} 
\vspace{0.2cm} 
\psfig{figure=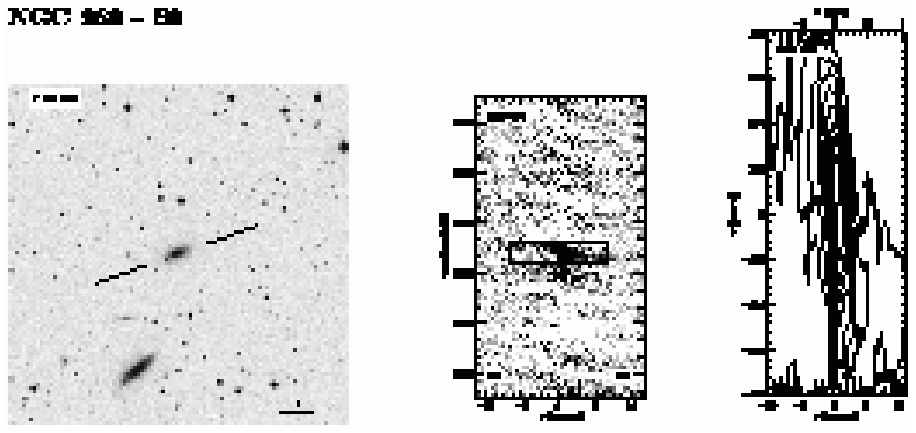,height=5.8cm} 
\vspace{0.2cm} 
\psfig{figure=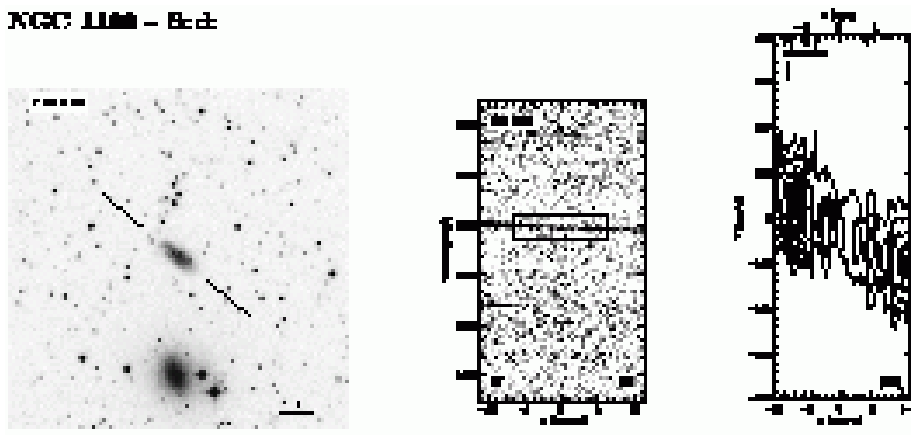,height=5.8cm} 
\vspace{0.2cm} 
\psfig{figure=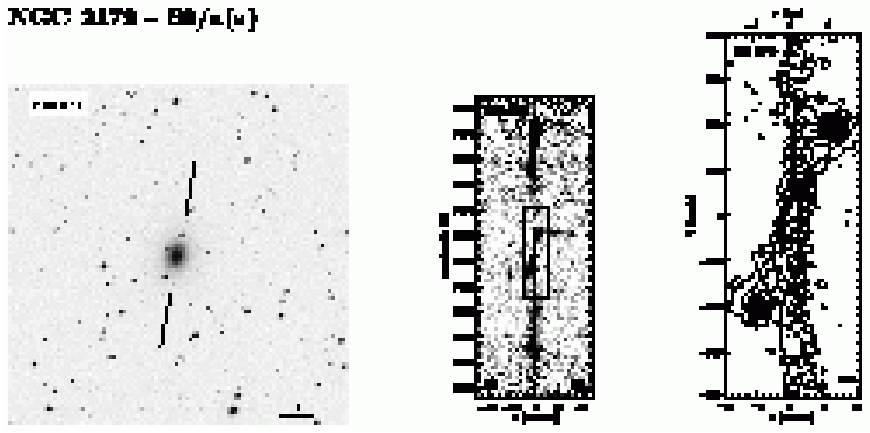,height=5.8cm} 
\caption{(continue)} 
\end{figure*} 
 
\addtocounter{figure}{-1} 
\begin{figure*} 
\psfig{figure=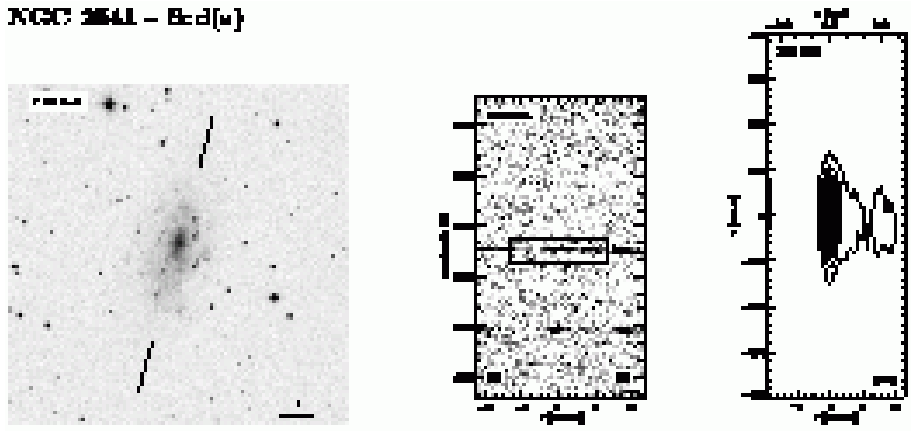,height=5.8cm} 
\vspace{0.2cm} 
\psfig{figure=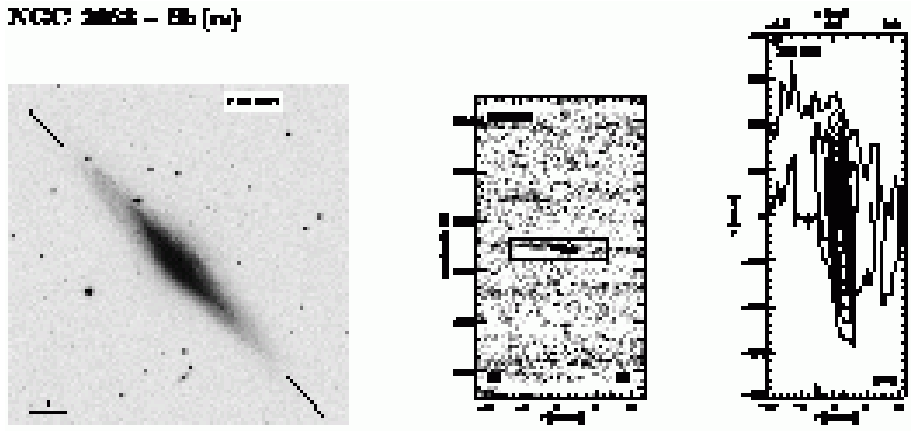,height=5.8cm} 
\vspace{0.2cm} 
\psfig{figure=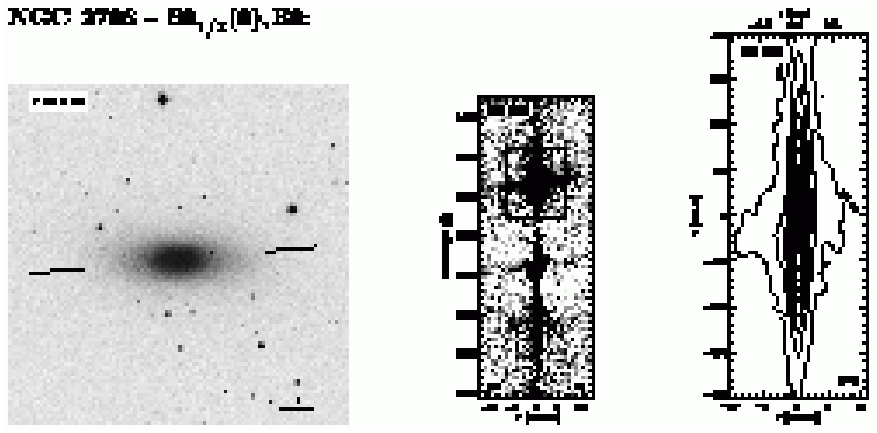,height=5.8cm} 
\vspace{0.2cm} 
\psfig{figure=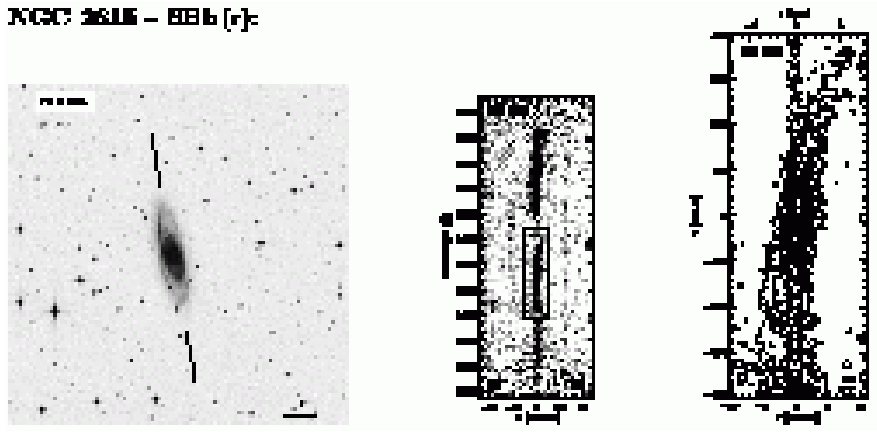,height=5.8cm} 
\caption{(continue)} 
\end{figure*} 
 
\addtocounter{figure}{-1} 
\begin{figure*} 
\psfig{figure=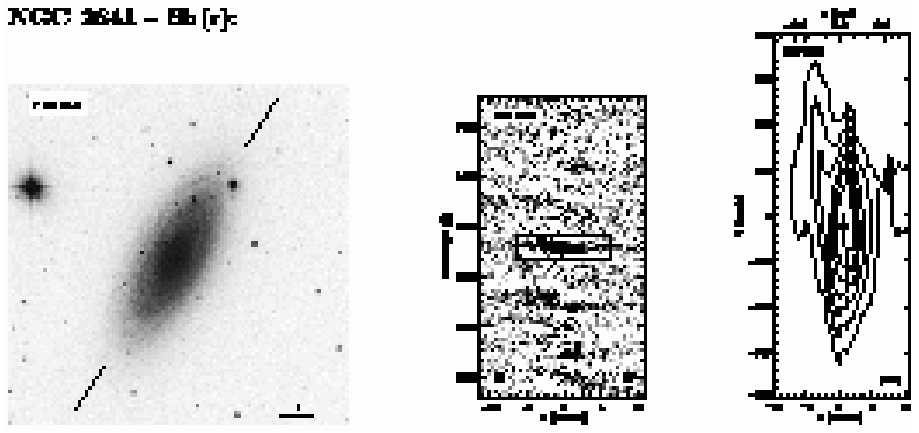,height=5.8cm} 
\vspace{0.2cm} 
\psfig{figure=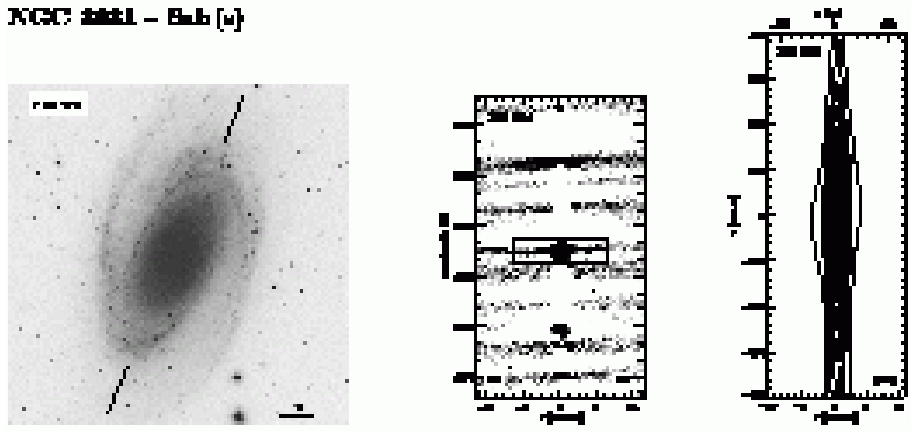,height=5.8cm} 
\vspace{0.2cm} 
\psfig{figure=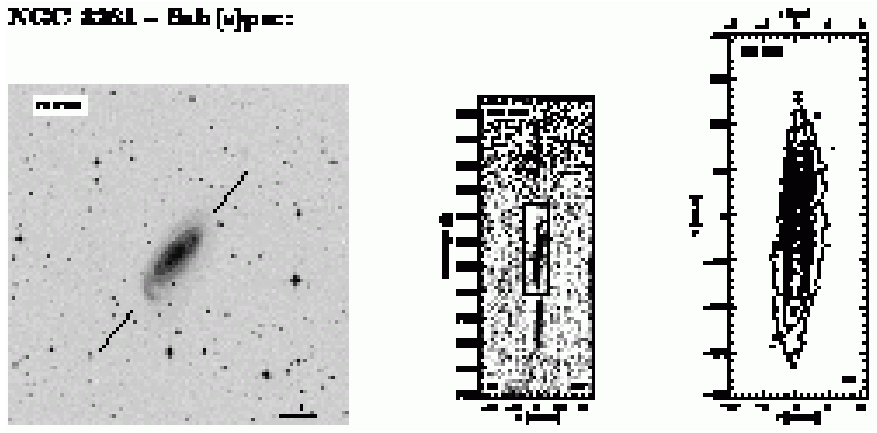,height=5.8cm} 
\vspace{0.2cm} 
\psfig{figure=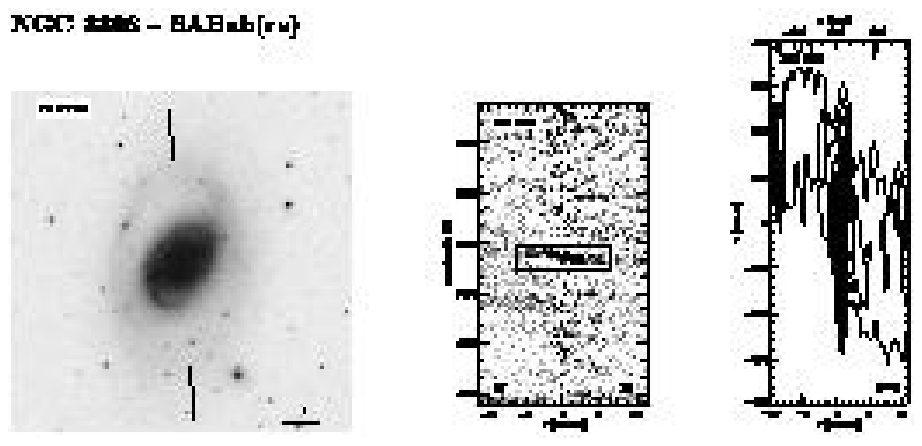,height=5.8cm} 
\caption{(continue)} 
\end{figure*} 
 
\addtocounter{figure}{-1} 
\begin{figure*} 
\psfig{figure=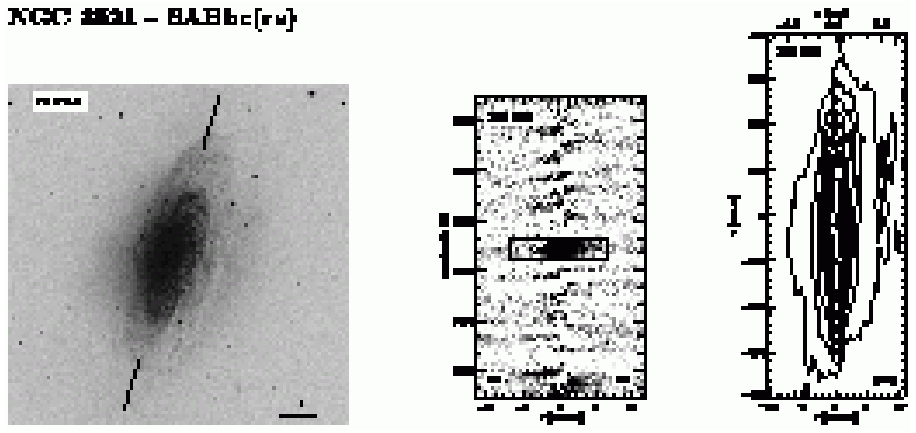,height=5.8cm} 
\vspace{0.2cm} 
\psfig{figure=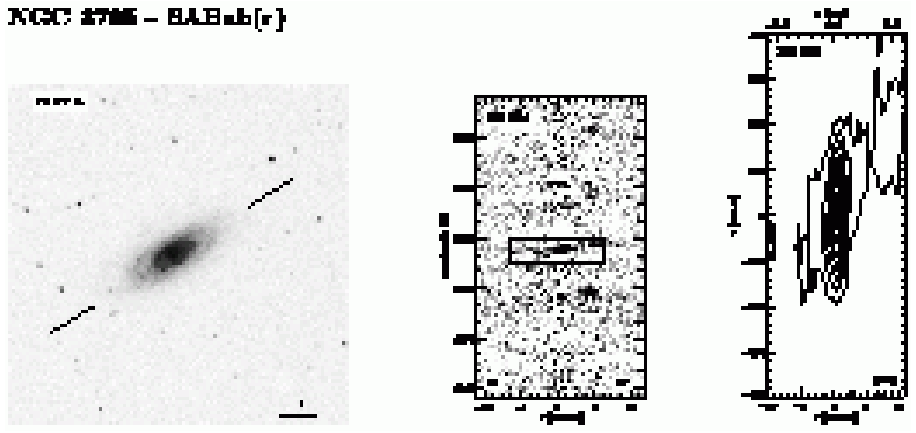,height=5.8cm} 
\vspace{0.2cm} 
\psfig{figure=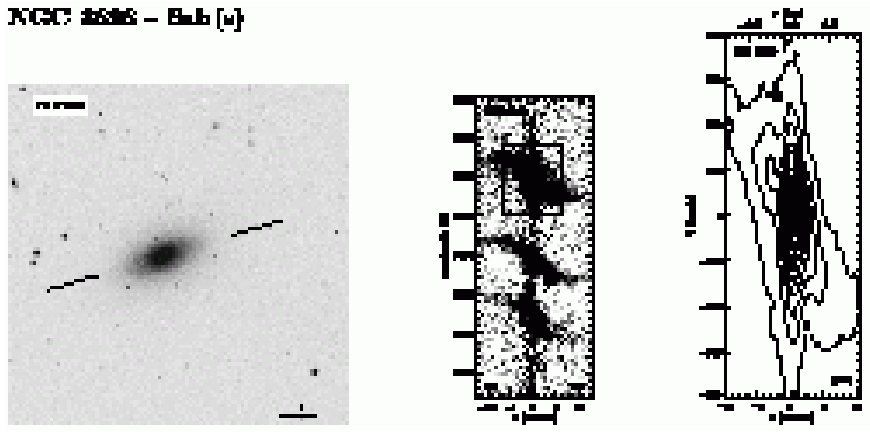,height=5.8cm} 
\vspace{0.2cm} 
\psfig{figure=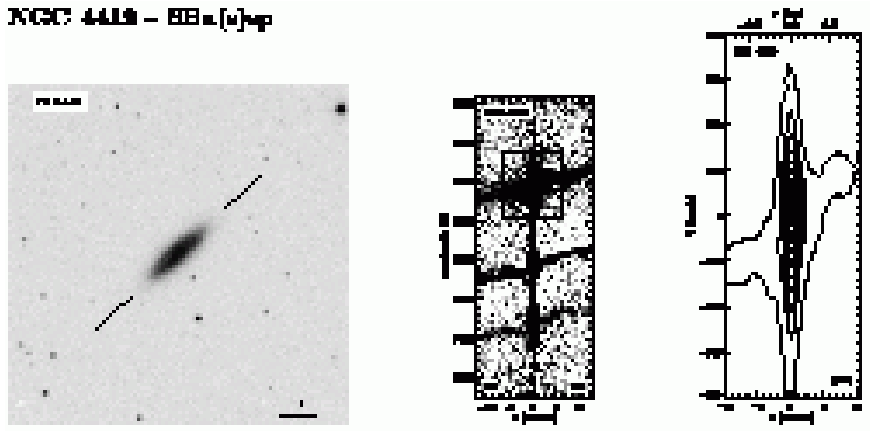,height=5.8cm} 
\caption{(continue)} 
\end{figure*} 
 
\addtocounter{figure}{-1} 
\begin{figure*} 
\psfig{figure=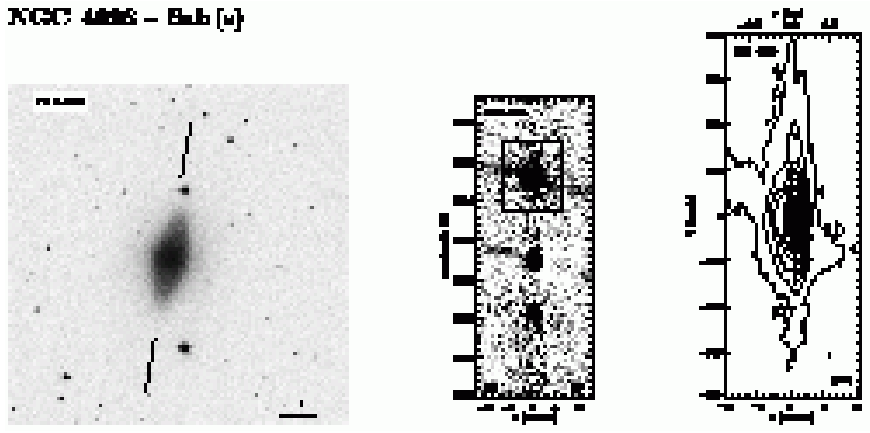,height=5.8cm} 
\vspace{0.2cm} 
\psfig{figure=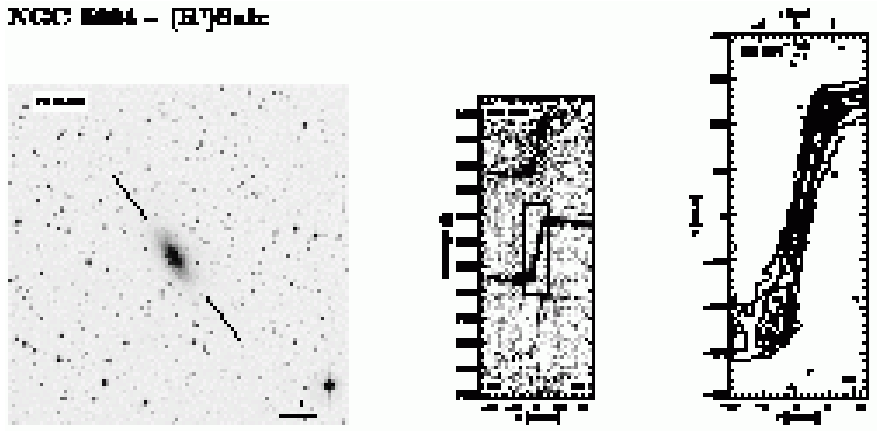,height=5.8cm} 
\vspace{0.2cm} 
\psfig{figure=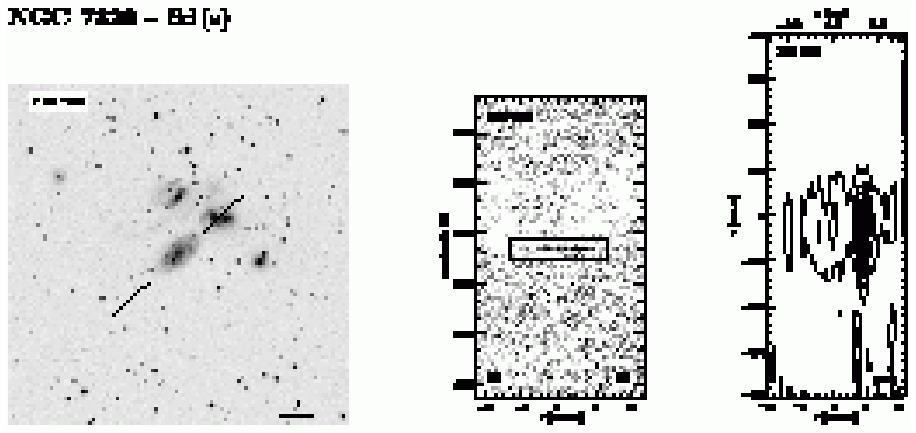,height=5.8cm} 
\vspace{0.2cm} 
\psfig{figure=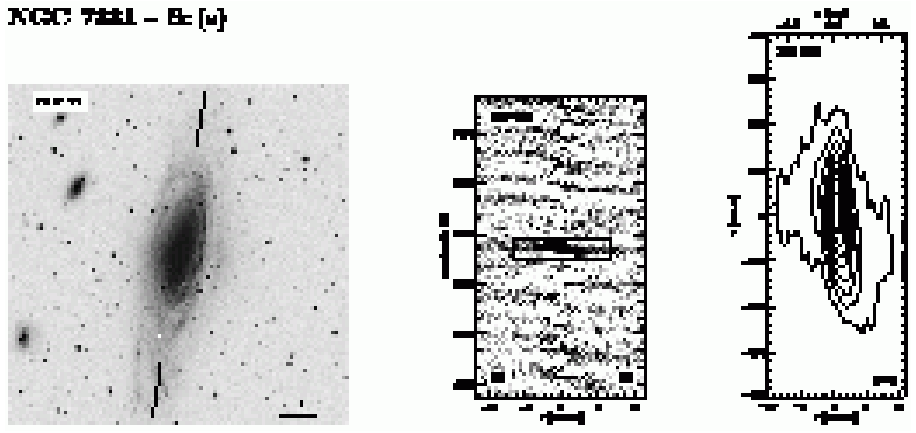,height=5.8cm} 
\caption{(continue)} 
\end{figure*} 
 
\addtocounter{figure}{-1} 
\begin{figure*} 
\psfig{figure=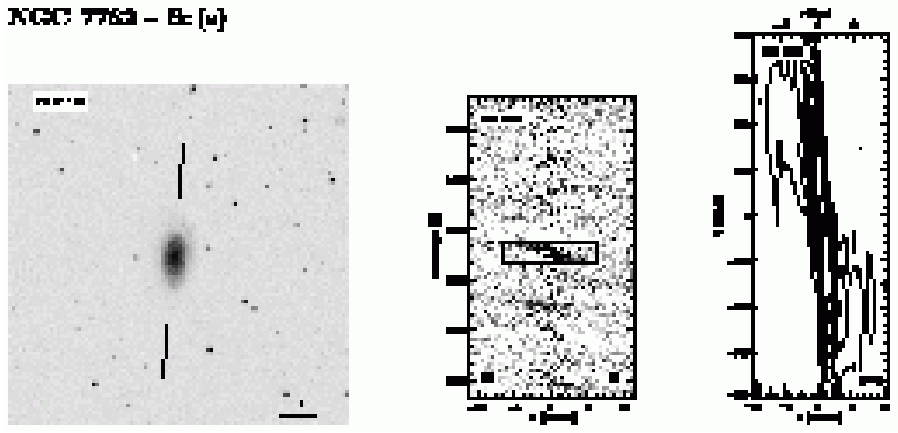,height=5.8cm} 
\caption{(continue)} 
\end{figure*} 
 
%\clearpage 
 
\begin{figure*} 
\centerline{ 
\hbox{ 
\psfig{figure=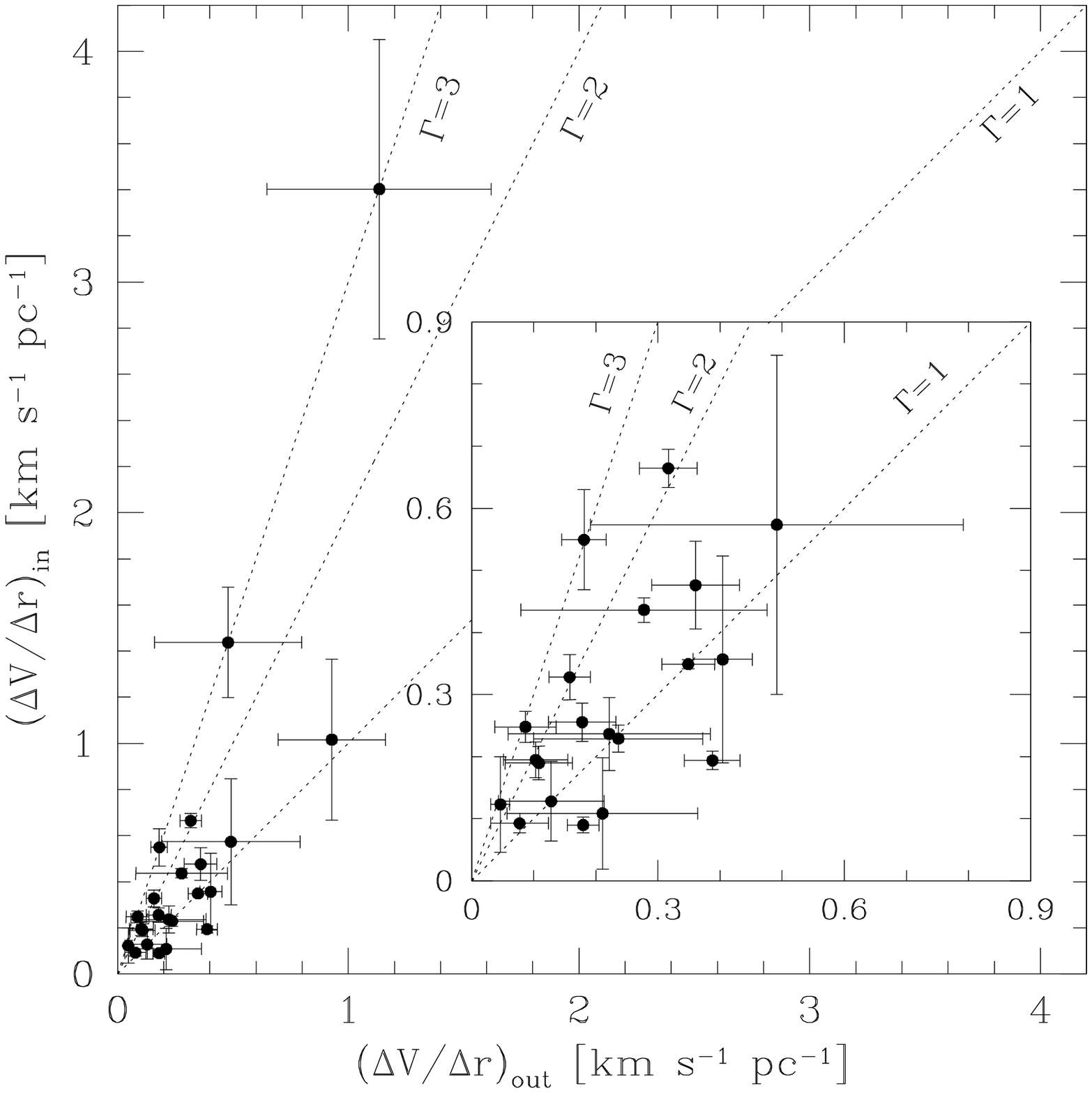,width=8.5cm} 
\psfig{figure=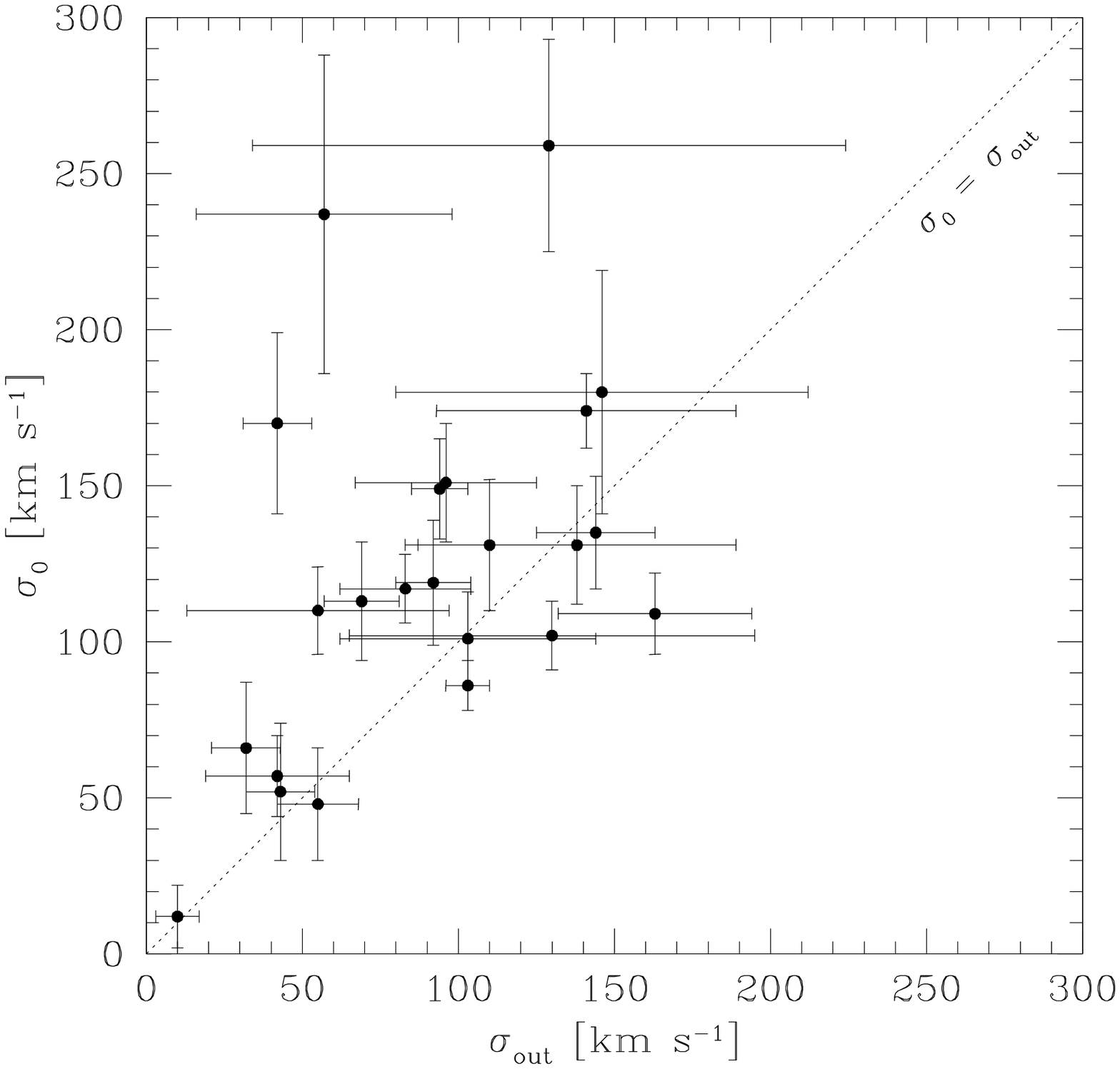,width=8.5cm} 
}} 
\caption{Inner vs. outer velocity gradients (left panel) 
  and central vs. outer velocity dispersions of the sample galaxies 
  (right panel). Observed velocity gradients collected in 
  Tab.~\ref{tab:measured_properties} have been corrected for galaxy 
  inclination and distance given in Tab.~\ref{tab:sample_properties}.} 
\label{fig:gradient_and_dispersion} 
\end{figure*}

\subsection{Measuring the position-velocity diagrams} 
\label{sec:measuring}

The PV diagrams plotted in Fig. \ref{fig:atlas} show the variety of
two-dimensional shapes of the optical emission lines in the inner
regions of disk galaxies. This finding is in agreement with the
earlier results of Rubin et al. (1997) and Sofue et al. (1998). We
suggest a classification of these PV diagrams based on the comparison
of the velocity gradients, velocity dispersions and fluxes measured at
different distances from the center.  Our goal is to identify galaxies
hosting a CNKD.
 
We measured the line-of-sight velocity, $V$, of the ionized gas at
$r\simeq\pm1''$ and $r\simeq\pm4''$ by fitting a Gaussian to the
relevant emission line.  The central wavelength of the Gaussian 
  fit was converted to velocity in the optical convention and
then the standard heliocentric correction was applied to obtain $V$.
The radii, $r\simeq\pm1''$, used for measuring the inner velocity
gradient \gradin , are dictated by the spatial resolution limit
imposed by seeing on our data ($\rm 0\farcs8 \la FWHM \la 1\farcs5$).
Choosing the smallest possible radii according to the seeing limit
assures us that if a central mass concentration is present, the
observed inner velocity gradient is maximized. The outer velocity
gradient, \gradout , measured at $r\simeq\pm4''$, serves as a
reference. In fact, for each sample galaxy the radius of influence,
$\theta_{\bullet}$, of the possible central mass concentration
predicted using the $M_{\bullet}-\sigma$ relation (Merritt \&
Ferrarese 2001a, see Tab.  \ref{tab:sample_properties}) is
$\theta_{\bullet} \ll 4''$. Therefore, \gradout\ is essentially
determined by the contribution of galaxy stellar component to the
potential.
 
We checked that velocity gradients did not significatively change if
the $\Delta V$ are computed from the difference of line-of-sight
velocity distribution maxima instead of the centers of the fitting
Gaussian fit. Moreover, to test the robustness of our measurements and
to estimate the associated uncertainties, we compute \gradin\ at
$r_{\it in} \simeq \pm0\farcs7\ {\rm and\ at}\ \pm1\farcs3$ 
and we compute \gradout\ at 
$r_{\it out} \simeq \pm3''\ {\rm and\ at}\ \pm5''$, respectively. 
The errors on \gradin\ have been estimated from the maximum difference
between the values measured at $\pm0\farcs7$ and $\pm1\farcs3$ with
respect to those measured at $\pm1''$.  Similarly, the errors on
\gradout\ have been estimated from the maximum difference between the
values obtained at $\pm3''$ and $\pm5''$ with respect to those
measured at $\pm4''$. The measured values of \gradin\ and \gradout\ 
are given in Tab. \ref{tab:measured_properties}.
 
The inner-to-outer velocity gradient ratio $\Gamma$ is independent of
galaxy inclination and has been adopted to indicate which galaxies are
characterized by rapidly-rotating gas in the inner regions. However,
to allow a direct comparison of their absolute values, we plotted the
inner velocity gradients as a function of the corresponding outer
velocity gradients in Fig. \ref{fig:gradient_and_dispersion}, after
correcting for galaxy inclination and distance given in Tab.
\ref{tab:sample_properties}. NGC~980, NGC~2179, NGC~2683, NGC~3031 and
NGC~7782 are clearcut cases of galaxies characterized by
$\Gamma\pm\Delta\Gamma>2$.
 
To characterize the velocity-dispersion and surface-brightness radial
profiles of each gaseous disk, we measured the velocity dispersion and
integrated flux of the ionized gas in the galaxy center and at
$r\simeq\pm4''$, using a Gaussian fit to the line adopted for the
velocity measurements.  The FWHM of Gaussian fit was corrected for
instrumental FWHM and converted into the velocity dispersion,
$\sigma$. The formal error of the fit has been adopted as the error on
the central value of velocity dispersion, while the errors on the
outer values have been estimated using the maximum difference between
the measurements obtained at $\pm3''$ and $\pm5''$ with respect to
those at $\pm4''$. The integrated flux was assumed to be the area of
the Gaussian fit and the associated error was estimated from Poisson
statistics. We considered only the central-to-outer integrated-flux
ratio since spectra were not flux calibrated. This process results in
line fluxes of different objects observed with different setups that
are not directly comparable.
The measured values of the velocity dispersion and the
central-to-outer integrated-flux ratio are given in Tab.
\ref{tab:measured_properties}. The central velocity dispersions are
shown as a function of the outer velocity dispersions in Fig.
\ref{fig:gradient_and_dispersion}. NGC~980, NGC~2179, and NGC~3031
exhibit the sharpest rises in observed velocity dispersion towards
their centers.

\subsection{A classification of the position-velocity diagrams} 
\label{sec:classification} 
 
We propose an operational classification of the PV diagrams of
Fig. \ref{fig:atlas} based on the properties we measured in Sect.
\ref{sec:measuring} and based on the analogy between the shapes of the
PV diagrams we observed and the shapes predicted for the spectrum of a
ionized thin gaseous disk, under the assumption that the gas moves in
circular orbits, in the plane of the galaxy. Although the model
assumptions may not be accurate in practice, the classification is
useful to identify the effects of rapid gas rotation. We do not
pretend to draw any general conclusions on the phenomenology of PV
diagrams from such a classification; it is just a tool adopted to
select the sample galaxies which possibly host CNKDs.
 
Under the model assumptions, two parameters are crucial in determining
the observed shape of the PV diagrams; they are the value of the
central mass concentration and the steepness of the intrinsic
surface-brightness distribution of the gaseous disk. To investigate
the change in the PV diagrams resulting from these two effects, we
have used the IDL modeling software developed in Bertola et al.\ 
(1998). We refer the reader to that paper for further details on the
model. A slit width and a seeing FWHM of 1\arcsec\ have been adopted,
with a spectrograph velocity scale of 10 km s$^{-1}$ pixel$^{-1}$ and
a spatial scale of $0\farcs3$ pixel$^{-1}$. The underlying galaxy
potential is assumed to result in a rigid rotation of 0.4 km
s$^{-1}$ pc$^{-1}$ in the plane of the disk, which is ``observed'' at
60$^\circ$ inclination ($i=90^\circ$ corresponding to edge-on). A
distance of 17 Mpc was adopted for the modeling, corresponding to the
distance of the Virgo cluster.
 
\begin{figure}[h!] 
\centerline{\hbox{ 
\psfig{figure=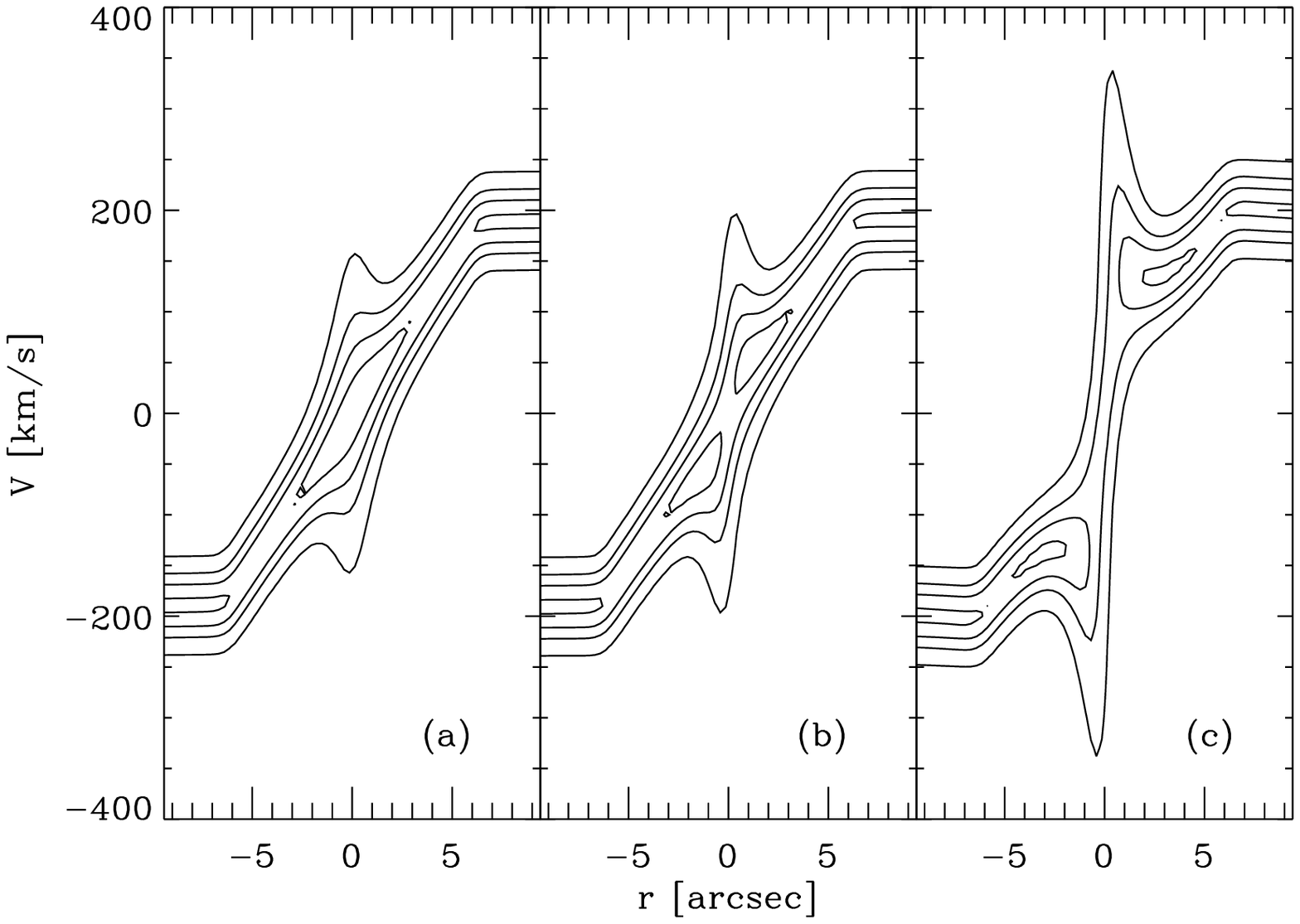,width=8.5cm} }} 
\centerline{\hbox{ 
\psfig{figure=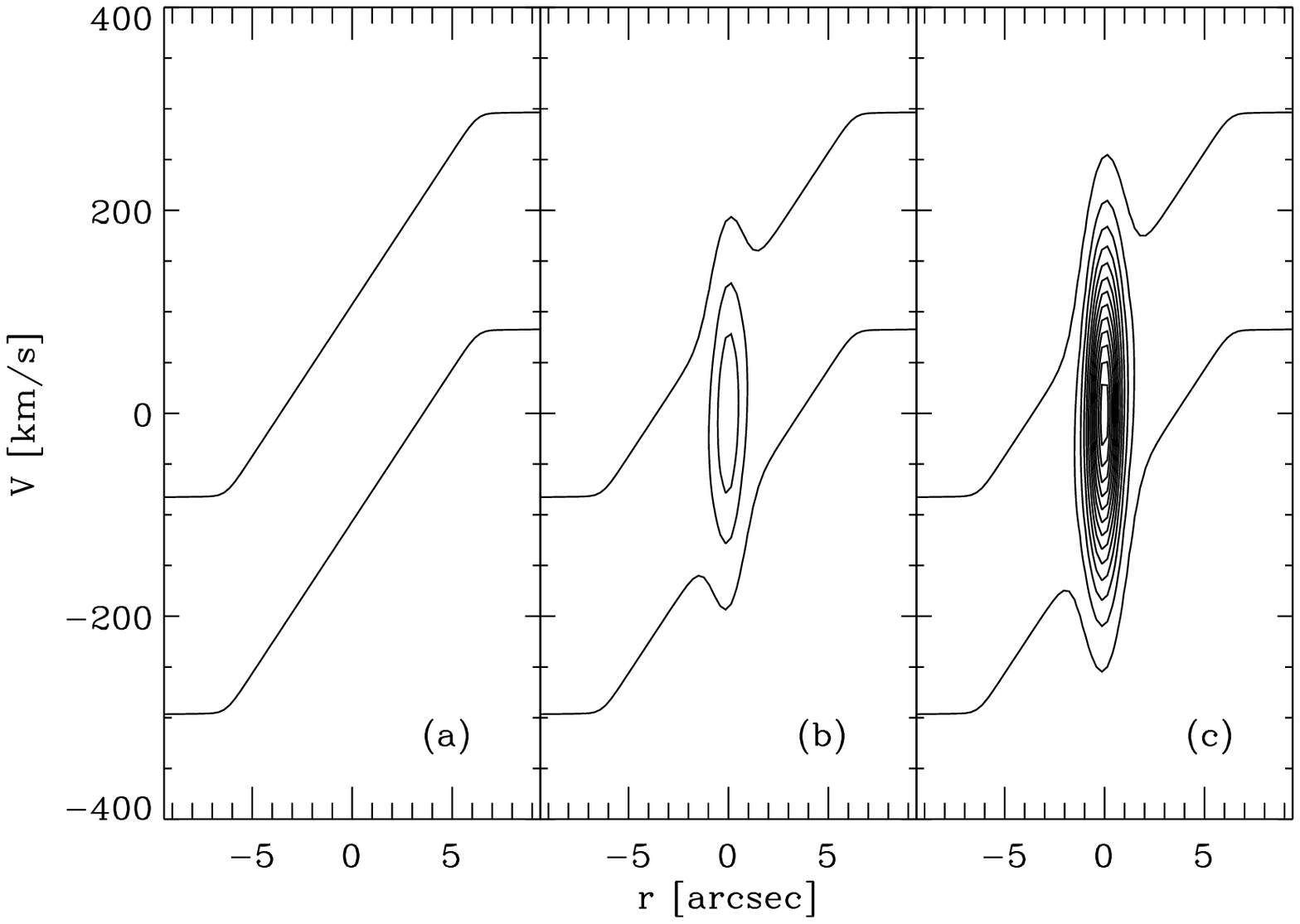,width=8.5cm} }} 
\caption{Upper panels: The shapes of PV diagrams as a function the  
  central mass concentration. The three panels represent the emission
  lines of CNKDs rotating around central pointlike sources of (a)
  $M_\bullet = 0$ \msun; (b) $M_\bullet = 10^8$ \msun; and (c)
  $M_\bullet = 10^9$ \msun.  The PV diagram in panel (c) is
  representative of Type I.
  Lower panels: The shape of PV diagrams as a function of the
  intrinsic surface brightness of the gas.  The three panels represent
  the emission lines of gaseous disks in rigid-body rotation with a
  projected velocity gradients ($\Delta V / \Delta r)_{\it
    in}$=($\Delta V / \Delta r)_{\it out}$=27 \kms\ arcsec$^{-1}$,
  observed velocity dispersions $\sigma_{0}=\sigma_{\it out}$=100 \kms
  , and nuclear fluxes (a) $F_0=F_{\it out}$; (b) $F_0=20\times F_{\it
    out}$; and (c) $F_0=100\times F_{\it out}$.  The PV diagrams in
  panels (a) and (c) are representative of Type II and III,
  respectively.}
\label{fig:modello} 
\end{figure}

The predicted effect on the PV diagram from an increase of the central
mass concentration is presented in the upper panels of
Fig.~\ref{fig:modello}. In this case, we assume an exponential the
surface-brightness profile superposed on a constant term:
$I(R)=I_0+I_1 \exp(-R/R_I)$, with $I_0=1$ and $I_1=5$ (in arbitrary
units) and $R_I=1\arcsec$, where the central mass is given by
$M_{\bullet}=0,10^8,10^9$ M$_\odot$ in panels (a), (b) and (c)
respectively.
 
The PV change that results from a variation in the brightness of a
central unresolved source is shown in the bottom panels of
Fig.~\ref{fig:modello}. In these panels the adopted surface-brightness
profile is assumed to be an essentially unresolved Gaussian superposed
on a constant term: $I(R)=I_0+I_1 \exp[-R^2/(2\sigma_I^2)]$, with
$I_0=1$, $\sigma_I=0\farcs3$ and $I_1=0,20,100$ (in arbitrary units)
in panel (a), (b) and (c) respectively.
 
\begin{figure} 
\centerline{ 
\hbox{ 
\psfig{figure=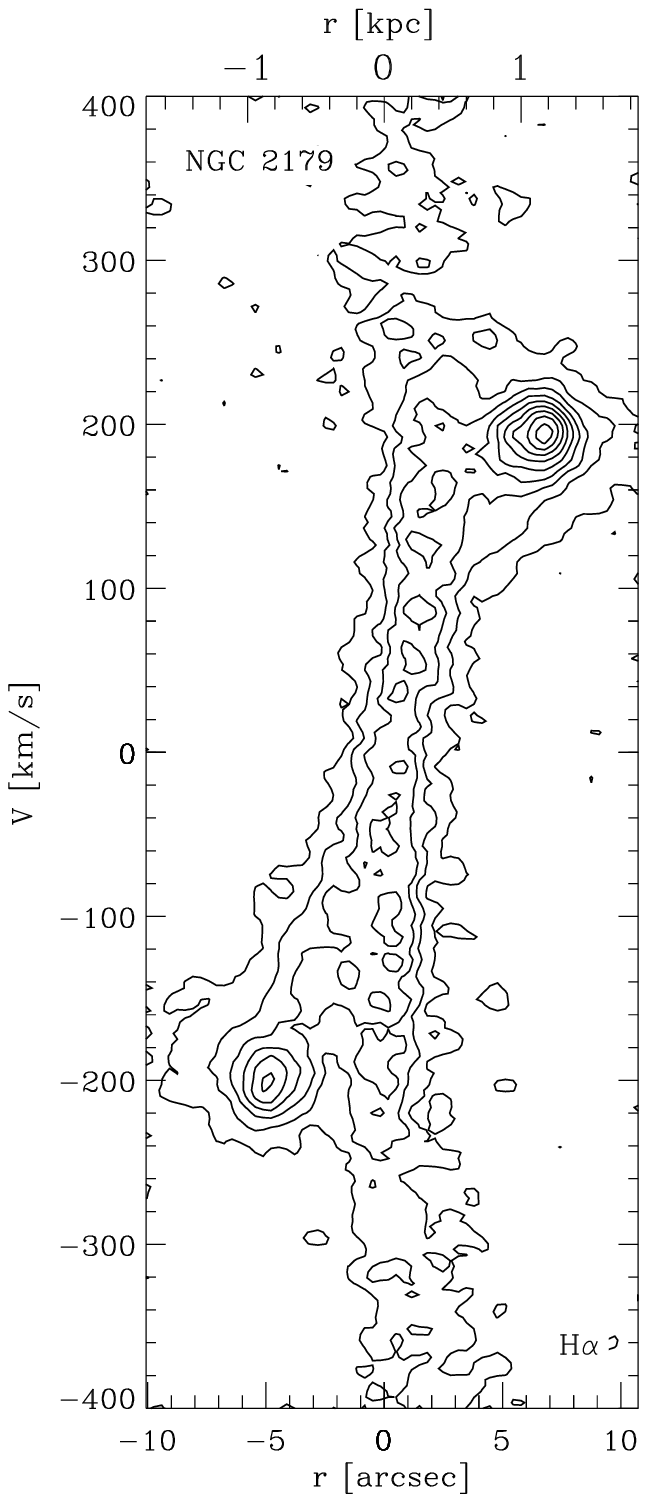,width=3cm} 
\psfig{figure=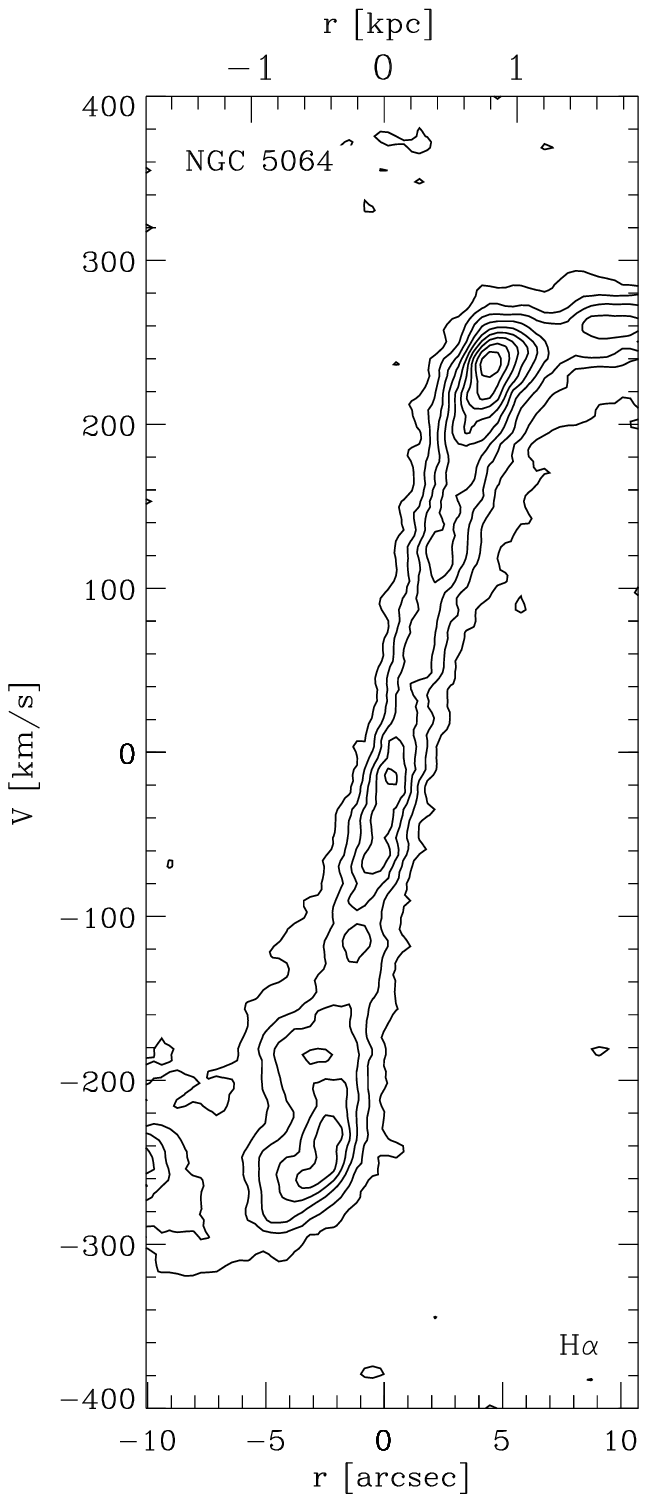,width=3cm} 
\psfig{figure=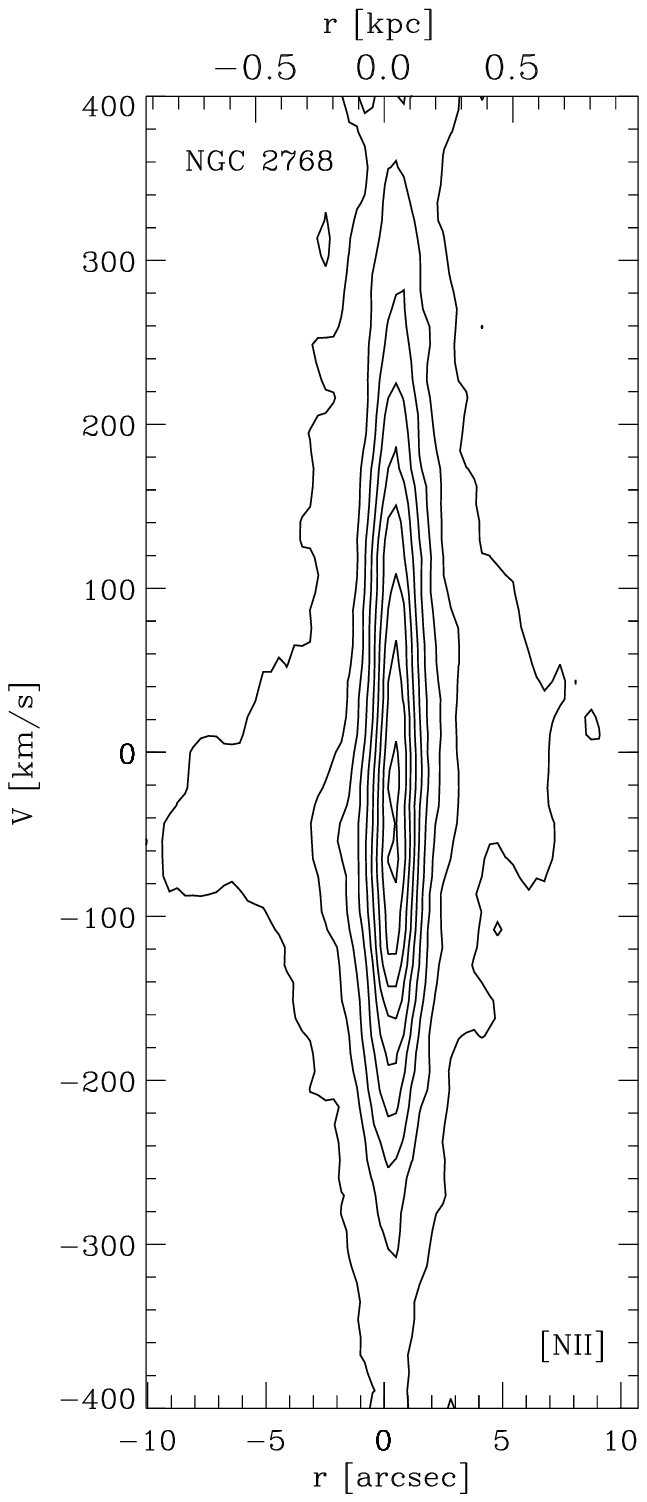,width=3cm} 
}} 
\caption{Contour plots of the prototypical examples of the three types 
  of PV diagrams. Left panel: NGC 2179 (Type I). Middle panel: NGC 
  5064 (Type II). Right panel: NGC 2768 (Type III).} 
\label{fig:examples} 
\end{figure}

By comparing the models of Fig.~\ref{fig:modello} with the observed PV 
diagrams, and inspecting the measured values of $\Gamma$, 
$\sigma_0/\sigma_{\it out}$ and $F_{0}/F_{\it out}$ we identify three 
different types of PV diagrams (see Fig. \ref{fig:examples}).

\medskip 
\noindent 
{\bf Type I.} This type of PV diagrams suggests the presence of two
distinct kinematical gaseous components. This results from the sharp
increase of \grad\ towards small radii, which indicates the presence
of a rapidly rotating gas in the innermost region of the galaxy.  The
inner-to-outer velocity gradient ratio is $\Gamma>2$ and the intensity
distribution along the line shows two symmetric peaks
with respect to the center.\\
The PV diagram of the Sa galaxy NGC~2179 (Fig. \ref{fig:examples}) can
be considered the prototype of this class. As we showed in Bertola et
al.\ (1998), the peculiar shape and intensity distribution of this PV
diagram can be modeled as a unique gaseous component that is rotating
in the combined potential of a central mass concentration embedded in
an extended stellar disk. Therefore the galaxies that exhibit this
kind of PV diagram (NGC 980, NGC 2179, and NGC 7782) are good
candidates to host a CNKD rotating around a central mass
concentration. They are ideal targets for HST spectroscopic follow-up
to constrain the mass of the possible SMBH. A good estimate of this
mass requires that the innermost kinematical points be within the
radius of influence (e.g.  Merritt \& Ferrarese 2001b). The three
galaxies meet this criterion, since the expected angular extension of
the radii of influence of their SMBHs are comparable to the pixel size
of the Space Telescope Imaging Spectrograph ($\theta_{\bullet} \approx
0\farcs05$).
An increase in the velocity dispersion ($\sigma_0 \ga 150$ \kms),
associated with a large increase in the velocity gradient (as in
NGC~980 and NGC~2179) is expected in the presence of a nuclear mass
concentration. It could result from differential motion within the
aperture or from intrinsic turbulence in the gaseous disk. On the
other hand, an increase in velocity dispersion that is {\it not\/}
associated with an increase in the velocity gradient may indicate that
the gas is {\it not\/} in a disk. However, a central mass
concentration may still be the cause of this increase in the velocity
dispersion.

\medskip 
\noindent 
{\bf Type II.} This class of PV diagram is characterized by a single 
velocity component which is in rigid-body rotation, as indicated by 
$\Gamma\approx1$. $\sigma_0/\sigma_{\it out}\approx1$ and $F_{0}/F_{\it 
out}\approx1$ are characteristic of these PV diagrams too.\\ 
We consider the PV diagram of the Sa galaxy NGC~5064 to be the
prototypical example of this kind of PV diagram (Fig.
\ref{fig:examples}).  In Bertola et al. (1998), we pointed out that in
this galaxy either the unresolved Keplerian part of the gaseous disk
does not result in a detectable contribution or the central mass
concentration is lower than $5\times10^7$ \msun .
Therefore we suggest that galaxies exhibiting this type of PV diagram
(see Tab. \ref{tab:measured_properties}) may harbor low-mass SMBHs,
although high spatial resolution spectroscopy and dynamical modelling
of the stellar kinematics are required to distinguish this possibility
from the effects of a peculiar gas distribution.

\medskip 
\noindent 
{\bf Type III.} PV diagrams of this type are characterized by an 
apparently broad nuclear emission-line component superimposed on a 
normal velocity curve. This results from the sharp increase of the line 
flux toward the center, as indicated by $F_{0}/F_{\it out}>1$.\\ 
The best example of this type of PV diagram is that of the S0 NGC~2768
(Fig.~\ref{fig:examples}). Most of the sample galaxies exhibit a PV
diagram belonging to this class because of a selection effect. They
have been observed because of their strong emission lines.

\medskip 
The classification and the peculiarities of the PV diagrams of all the
sample galaxies are discussed in the appendix.

\section{Discussion and conclusions} 
\label{sec:conclusions} 
 
In Bertola et al. (1998) we demonstrated that it is possible to detect
the signature of a CNKD in the emission-line PV diagrams obtained from
ground-based spectroscopy of nearby galaxies. In addition, we showed
that once properly modelled, this technique can lead to reliable
upper-limits for the central mass concentration.
Using these results, and to identify new galaxies hosting a CNKD, we
obtained long-slit spectra along the major axes of 23 nearby disk
galaxies, measuring the PV diagrams of the ionized-gas components from
the emission lines.
 
To achieve our goal, we analyzed these emission-line PV diagrams by
measuring the values of the velocity gradient, the velocity
dispersion, and the integrated flux of the ionized gas at different
radii, and by comparing the shape of the observed emission lines with
that predicted for a thin disk of gas moving in circular orbits in the
galaxy plane. This classification allowed us to recognize the possible
presence of a CNKD in 3 of our objects. Recently, Rubin et al. (1997)
and Sofue et al. (1998) discussed the kinematics of rapidly-rotating
gas disks observed in the central few hundred parsecs of S0's and
spiral galaxies. By combining our sample with their samples, we find
that the PV diagrams of 9 out 54 galaxies exhibit a $\Gamma>2$, which
is indicative of a CNKD. The fact that in the majority of these cases
CNKDs have not been observed means that not all of these galaxies
enough gas detectable and rotating Keplerian orbits. Therefore, we
estimated that the frequency of CNKDs, measured from ground-based
spectroscopy of emission-line disk galaxies, is $\la20\%$.
This result is consistent with the findings of Sarzi et al. (2001),
which are based on HST spectroscopy. Indeed, they found a gaseous disk
with a well-ordered velocity field suitable for dynamical modelling at
HST resolution in only 4 of the 23 randomly selected emission-line
disk galaxies they observed.
 
The analysis of ground-based PV diagrams allows identification of
those galaxies that are good candidates for hosting a CNKD rotating
around a central mass concentration, and are therefore are good
candidates for follow-up HST spectroscopy. It is worth noting that in
this way it is possible to improve the present low success rate of HST
programs aimed at estimating SMBH masses in nearby bulges by modelling
nuclear gas kinematics.
Although the SMBH mass hosted by the candidates selected by
ground-based observations are expected to lie in the high-mass end of
the M$_{\bullet}-\sigma$ diagram because of the resolution limits
imposed on their PV diagrams by seeing, these candidates are required
to elucidate the relationship for disk galaxies, which are
underrepresented relative to elliptical galaxies in the sample of
galaxies studied so far.

\acknowledgements 
We thank Betsy Gillespie for reading the manuscript.
This research has made use of the Lyon-Meudon Extragalactic Database
(LEDA) and Digitized Sky Survey (DSS).

\appendix 
\section{Notes on individual galaxies} 
\label{sec:appendix} 
 
The PV diagrams of the sample galaxies are plotted in Fig.
\ref{fig:atlas} and their classifications according to the criteria
proposed in Sect. \ref{sec:classification} are given in Tab.
\ref{tab:measured_properties}.  In this section, we describe the most
important properties of the sample galaxies and discuss the
classification of their PV diagrams. The values of the velocities and
the velocity dispersions include no corrections for inclination.

\medskip 
\noindent 
{\bf NGC~470.} This intermediate-type spiral hosts either two 
nested bars (Wozniak et al. 1995) or a bar with a triaxial bulge 
(Friedli et al. 1996). 
The velocity gradient and velocity dispersion of the \og\ line do not
significantly change moving from the nuclear to the outer regions,
while its integrated flux increases towards the center. We
classify the PV diagram of NGC~470 as Type III.

\medskip 
\noindent 
{\bf NGC~772.}  According to the dynamical modelling of Pignatelli et
al. (2001), the velocity of the ionized gas traces the circular speed
in gravitational equilibrium.  The PV diagram of NGC 772 derived from
the \og\ emission line is classified as Type III since the integrated
flux of the emission line in the nucleus is greater than that measured
in the outer regions.

\medskip 
\noindent 
{\bf NGC~949.}  As in the previous case, also the Type III
classification of the PV diagram of NGC~949 results from the increase
of the integrated flux of the \og\ line toward the center.

\medskip 
\noindent 
{\bf NGC~980.} In the PV diagram of this S0 galaxy, a tilted and
bright component appears to be superimposed on a normal rotation
curve. The increase of both the velocity gradient and the velocity
dispersion toward the center are indicative of the kinematics of a
CNKD and give this PV diagram its Type I classification.  This is also
the case for NGC 2179 and NGC 7782.
 
\medskip 
\noindent 
{\bf NGC~1160.}  The PV diagram of this Scd spiral is characterized by
a constant velocity gradient and a constant integrated flux for the
\og\ line.  This makes its Type II classification straightforward.

\medskip 
\noindent 
{\bf NGC~2179.}  The PV diagram of NGC 2179 is the prototype of the
Type I class.
The two-dimensional shapes of the emission lines are such that they
gives the erroneous impression of two distinct velocity components.
One apparent velocity component has the form of a highly tilted
straight line rising from zero velocity in the galaxy center (i.e.,
a faster-rising rotation curve).  The other component is a less-tilted
straight line (i.e., a slower-rising rotation curve) superimposed on
the first.  Both lines naively appear to imply solid-body rotation in
the inner parts of this galaxy; both lines culminate as the radius
increases to almost the same maximum velocity; the slower-rising
rotation curve shows a flat portion in its outer regions.  Rather than
being of two different physical origins, we have shown that properly
accounting for the seeing, slit width, and pixel size effects, these
two apparently solid-body rotation curves can be modelled as the
velocity field of a thin gaseous disk rotating in the combined
gravitational potential of central point-like mass and an extended
stellar disk (see Bertola et al. 1998 for details).

\medskip 
\noindent 
{\bf NGC~2541.} The bright central component in this PV diagram is
characterized by the same velocity dispersion measured in the outer
parts of the disk. This is typical of Type III diagrams.

\medskip 
\noindent 
{\bf NGC~2683.} {According to Merrifield \& Kuijken (1999), the 
PV diagram of NGC 2683 has a `figure-of-eight' shape produced by the 
presence of two kinematically distinct gaseous components. 
This feature is barely visible in our PV diagram because of the lower
S/N ratio of the spectrum. Although the properties of the PV diagram
of NGC 2683 are similar to those of NGC~980, NGC~2179 and
NGC~7782, it does not warrant a Type I classification.
Indeed in NGC 2683 we are observing two gas components which are
spatially distinct and superimposed along the line of sight because of
the high inclination of the galaxy ($i=78\dg$). They are generated by
the presence of a non-axisymmetric potential (Kuijken \& Merrifield
1995; Bureau \& Athanassoula 1999).
 
This is not the case for NGC 980, NGC 2179 and NGC 7782, which are
less inclined ($i=58\dg,47\dg$ and $58\dg$, respectively) and
unbarred, exhibiting unique gaseous components.

\medskip 
\noindent 
{\bf NGC~2768.} The presence of a definite outer envelope with subtle
dust patches surrounding the bulge (see panels 38 and 53 in CAG)
supports the S0 classification of this galaxy, which appears as an
E6$\ast$ in RC3. The kinematical decoupling between the ionized gas
and the stars, detected by Bertola, Buson \& Zeilinger (1992), has
been interpreted as a result of the presence of gas orbiting in a
polar ring (M\"ollenhoff, Hummel \& Bender 1992; Fried \& Illinghworth
1994).
The inner velocity gradient is higher than the outer one which is one
of the lowest we measured (\gradout$ = 0.09$ \kmspc ). The constant
velocity dispersion and the steep increase of \ng\ flux in the center
imply a Type III classification for this PV diagram.

\medskip 
\noindent 
{\bf NGC~2815.} The presence of the spectrum of broad emission lines
in the nuclear region of NGC~2815 with wings which are very close each
to other makes the subtraction of the galaxy continuum critical.
Indeed, residual continuum is still visible in the PV diagram derived
from the \ha\ line; it gives the erroneous impression that the PV
diagram shape is similar to that of NGC~2179. We classify this PV
diagram as Type III class because of its constant velocity gradient
and large central-to-outer integrated-flux ratio.

\medskip 
\noindent 
{\bf NGC~2841.} According to Sil'Chenko, Vlasyuk \& Burenkov (1997)
the ionized gas is rotating orthogonally with respect to the galaxy
plane in the inner $5''$. Alternatively, Sofue et al.  (1998) reported
that the central portion of the PV diagram derived from the \ha\ and
\ng\ lines is slightly tilted in the direction of the galactic
rotation, suggesting the presence of a rapidly rotating nuclear disk.
Our PV diagram exhibits a complex and asymmetric shape that could be
related to these different kinematic components. However, we do not
measure a significant variation of the velocity gradient or the
velocity dispersion with radius. The line flux increases slightly
toward the center. These features are similar to those of PV diagrams
included in the Type III class, and make it difficult to associate
with the central component to a fast-rotating disk as by indicated
Sofue et al. (1998).

\medskip 
\noindent
{\bf NGC~3031.}  HST \ha\ imaging reveals the presence of a nuclear
gaseous disk (Dereveux, Ford \& Jacoby 1997) similar in size and shape
to the CNKD of M87 (see Macchetto et al. 1997 and references
therein). The disk is rotating around a SMBH with $M_{\bullet} =
3\times10^6$ \msun , according to determinations based on stellar
kinematics (Bower et al.  1996) and broad-line emission (Ho,
Filippenko \& Sargent 1996).
The spatial and spectral resolution of our spectrum allow us only to
detect the presence of a broad and bright central component in the PV
diagram. In fact, it exhibits the highest central-to-outer
integrated-flux ratio of our sample, which warrants a Type III
classification.  From the available spectrum, it is difficult to claim
that NGC 3031 is hosting a CNKD even though we measure a remarkably
large \gradin\ ($= 3.4$ \kmspc ) and a large inner-to-outer
velocity-gradient ratio ($\Gamma=3.0$).

\medskip 
\noindent 
{\bf NGC~3281.} The ionized-gas kinematics measured by Rubin et al. (1985) 
and Corsini et al. (1999) extends out to about $50''$ from the nucleus 
but in our spectrum the emission is confined in the innermost $5''$. 
The inner velocity gradient is steeper than the outer one and this
early-type spiral has one of the highest central-to-outer flux ratios
of the whole sample. Even if the emission is not extended we consider
the PV diagram of NGC~3281 to be of Type III because of its intense
nuclear emission.

\medskip 
\noindent 
{\bf NGC~3368.} From NIR photometry, Jungwiert, Combes \& Axon (1997)
identified a possible double-barred structure within this Sab spiral.
Although the PV diagram seems to have a two-component structure, the
constant velocity gradient and steep increase of the integrated flux
of the \og\ line toward the center suggest a Type III classification.

\medskip 
\noindent 
{\bf NGC~3521.} The PV diagram of this intermediate-type spiral has 
been recently measured by Sofue et al. (1998) from the \ha\ and the 
\ng\ emission lines. They interpreted the central component  
observed in the \ng\ line as an indication of the presence of a fast  
rotating gaseous disk in the nucleus. 
We suggest that this feature, which is clearly visible also in the PV
diagram we derived from the \og\ emission line, results from the
increase of the line flux rather than the velocity gradient.  Morever,
the velocity dispersion does not change with radius. Therefore it is
Type III diagram.

\medskip 
\noindent 
{\bf NGC~3705.} The Type III classification of this PV diagram mostly
results from the centrally-peaked radial profile of the integrated
flux of the \og\ line, which gives the impression of a steep central
component superimposed on a slowly-rotating component.

\medskip 
\noindent 
{\bf NGC~3898.} The ionized-gas distribution and kinematics of this Sa
galaxy have recently been studied in detail by Pignatelli et al.
(2001). They found that in the innermost region ($|r|\la8''$) of NGC
3898, the ionized gas is rotating more slowly than the circular
velocity predicted by dynamical modelling based on stellar kinematics
and photometry.
The fingerprint of such a `slowly-rising' rotation curve (according to
Kent 1988 definition) can be recognized in the decrease of the
velocity gradient at smaller radii.  The two-component shape of the PV
diagram results from the bright nuclear emission and not the increase
of velocity gradient or the velocity dispersion. This is a Type III PV
diagram and its similarity to the PV diagram of NGC 4419 is
remarkable.

\medskip 
\noindent 
{\bf NGC~4419.}  The spectrum and consequently the PV diagram of NGC
4419 are similar to those of NGC 3898. The spectra show the same
strong and broad \ha\ absorption and the PV diagrams are characterized
by the same bright central component. They have also similar
inner-to-outer velocity gradient and dispersion ratios and both belong
to the Type III class.
As NGC 3898 also NGC 4419 is one of the bulge-dominated spirals
displaying a slowly-rising rotation curve of the ionized gas discussed
by Kent (1988).

\medskip 
\noindent 
{\bf NGC~4698.}  This Sa galaxy shows a remarkable orthogonal
geometrical and kinematical decoupling between the inner portion of
the bulge and galaxy disk (Bertola et al. 1999).
The asymmetric shapes of the \ha\ and \nii\ lines are seen at a simple
visual inspection of the spectrum, and they are more evident in the PV
diagram obtained from the \ng\ line.  Although we measured an increase
of the velocity gradient towards the center, we note that NGC~4698 has
the shallowest outer gradient of all the sample galaxies (\gradout$ =
0.05$ \kmspc ). This gradient corresponds to the central plateau
measured in the ionized-gas rotation curve by Bertola \& Corsini
(2000). The Type III classification has been assigned to this PV
diagram on the basis of its high central-to-outer integrated-flux
ratio.

\medskip 
\noindent 
{\bf NGC~5064.}  The PV diagram of NGC~5064 is the prototype of the
Type II class.
It is useful to compare the emission-line spectrum of NGC~5064 to that
of NGC~2179, because both spectra have been obtained with same setup
and observing conditions. In contrast with NGC~2179, the emission-line
spectrum of NGC~5064 does not show any peculiar features; there is
only one component in the central region. The velocity increases
linearly with radius until it reaches about $200$ \kms , $4''$ from
the center.  The velocity dispersion and the integrated flux of the
\ha\ line remain almost constant in this radial range.

\medskip 
\noindent 
{\bf NGC~7320.} The spectrum we obtained for this late-type spiral,
which belongs to Stephan's Quintet is of poor quality. The \og\ line
shows a bright knot at about $5''$ from the center resulting in the
observed $F_{0}/F_{\it out}=0.3$. The Type II classification of the PV
diagram is based on the constant inner-to-outer velocity gradient and
velocity dispersion.

\medskip 
\noindent 
{\bf NGC~7331.} The presence of a SMBH ($M_\bullet \sim 10^8$ \msun)
in the center of NGC 7331 has been debated by different authors
(Afanasiev, Morozov \& Levi 1989; Bower et al. 1993; Mediavilla et al.
1997; Sil'Chenko 1999). The debate centers on observations of the
distribution and kinematics of ionized gas.
Our PV diagram is similar to that of NGC 772. We measure an increase
of the integrated flux of the \og\ line at smaller radii, along with
constant velocity gradient and to slight increase of the velocity
dispersion. The PV diagram of NGC~7331 is of Type III.

\medskip 
\noindent 
{\bf NGC~7782.}  The inner region of this PV diagram is characterized
by a sharp increase of the velocity gradient, as confirmed by the
large inner-to-outer velocity-gradient ratio we measure. The \og\ line
exhibits a bright nuclear component and it velocity dispersion rapidly
decreases with radius. The properties of the PV diagram of NGC~7782
are close to those of NGC 2179 leading to the Type I
classification. The nuclear ionized-gas kinematics of NGC 7782 is
indicative of a CNKD.

\end{document}